\documentclass[reprint,prd, superscriptaddress, tightenlines, longbibliography, nofootinbib, eqsecnum, amsfonts, amsmath, floatfix, notitlepage]{revtex4-1}
\pdfoutput=1

\usepackage[utf8]{inputenc}
\usepackage{mathrsfs}
\usepackage{euscript}
\usepackage{epsfig}
\usepackage{graphics}
\usepackage{graphicx}
\usepackage{comment}
\usepackage{amsmath}
\usepackage{amssymb}
\usepackage{bm}
\usepackage[usenames,dvipsnames,svgnames,table]{xcolor}
\usepackage{xspace}
\usepackage{wasysym}
\usepackage{times}
\usepackage{appendix}
\usepackage{lipsum}
\usepackage[nolist,nohyperlinks]{acronym}
\usepackage{float}
\usepackage{simplewick}
\usepackage{natbib, ifthen}
\usepackage{hyperref}
\usepackage{longtable}

\defcitealias{Ade:2015zua}{PL2015}
\renewcommand{\vr}{\boldvec{r}}
\newcommand{\PlanckLens}{\citetalias{Ade:2015zua}}
\newcommand{\CAMB}{{\sc camb}}

\newcommand{\orcidacronym}[2]{\href{http://orcid.org/#2}{#1}}
\newcommand{\ALorcid}{\orcidacronym{AL}{0000-0001-5927-6667}}
\newcommand{\JCorcid}{\orcidacronym{JC}{0000-0002-5751-1392}}

\newcommand{\la}{\langle}
\newcommand{\ra}{\rangle}

\newcommand{\mksym}[1]{\ifmmode {\rm #1}\else #1\fi}

\providecommand{\Planck}{\textit{Planck}}
\providecommand{\planck}{\Planck}

\providecommand{\text}[1]{\rm{#1}}

\providecommand{\CAMB}{{\tt camb}}

\newcommand{\begm}{\begin{pmatrix}}
\newcommand{\enm}{\end{pmatrix}}

\newcommand\ba{\begin{eqnarray}}
\newcommand\ea{\end{eqnarray}}
\newcommand\bea{\begin{eqnarray}}
\newcommand\eea{\end{eqnarray}}

\newcommand\be{\begin{equation}}
\newcommand\ee{\end{equation}}

\newcommand{\vell}{{\boldsymbol{\ell}}}

\newcommand{\ud}{{\rm d}}

\newcommand{\boldvec}[1]{{\mbox{\boldmath{$#1$}}}}

\providecommand{\vr}{\boldvec{r}}

\newcommand{\vx}{\boldvec{x}}
\newcommand{\vy}{\boldvec{y}}
\newcommand{\vz}{\boldvec{z}}

\newcommand{\clw}{\mathcal{W}}

\newcommand{\isdraft}[1]{}

\newcommand{\AL}[1]{{\isdraft{\color{blue} AL: #1}}}

\renewcommand{\AC}[1]{{\isdraft{\color{green} AC: #1}}}

\newcommand{\del}{\textrm{del}}
\newcommand{\dat}{\textrm{dat}}
\newcommand{\lag}{\left \langle}
\newcommand{\rag}{\right \rangle}
\newcommand{\av}[1]{\lag #1 \rag}
\newcommand{\deflect}{\boldsymbol{\alpha}}
\newcommand{\invdeflect}{\boldsymbol{ \beta}}
\newcommand{\hn}{{ \boldsymbol{n}}}
\newcommand{\debiasedChat}{\hat C_{\ell,\rm{debias}}^{\rm del}}

\newcommand{\revision}[1]{#1}

\begin{document}

\newcommand{\Sussex}{Department of Physics \& Astronomy, University of Sussex, Brighton BN1 9QH, UK}


\title{Internal delensing of Planck CMB temperature and polarization}
\author{Julien Carron}\email{j.carron@sussex.ac.uk}
\affiliation{\Sussex}
\author{Antony Lewis}
\affiliation{\Sussex}
\homepage{http://cosmologist.info}
\author{Anthony Challinor}
\address{Institute of Astronomy and Kavli Institute for Cosmology, Madingley Road, Cambridge, CB3 0HA, UK}
\address{DAMTP, Centre for Mathematical Sciences, University of Cambridge, Wilberforce Road, Cambridge CB3 OWA, UK}

\begin{abstract}
We present a first internal delensing of CMB maps, both in temperature and polarization, using the public foreground-cleaned (SMICA) Planck 2015 maps.
After forming quadratic estimates of the lensing potential, we use the corresponding displacement field to undo the lensing on the same data. We build differences of the delensed spectra to the original data spectra specifically to look for delensing signatures.
After taking into account reconstruction noise biases in the delensed spectra, we find an expected sharpening of the power spectrum acoustic peaks with a delensing efficiency of 29\,\% ($TT$) 25\,\% ($TE$) and 22\,\% ($EE$). The detection significance of the delensing effects is very high in all spectra: $12\,\sigma$ in $EE$ polarization; $18\,\sigma$ in $TE$; and $20\,\sigma$ in $TT$. The null hypothesis of no lensing in the maps is rejected at $26\,\sigma$.  While direct detection of the power in lensing $B$-modes themselves is not possible at high significance at Planck noise levels, we do detect (at $4.5\,\sigma$ \revision{under the null hypothesis}) delensing effects in the $B$-mode map, with $7\,\%$ reduction in lensing power. Our results provide a first demonstration of polarization delensing, and generally of internal CMB delensing, and stand in agreement with the baseline $\Lambda$CDM Planck 2015 cosmology expectations.
\end{abstract}

\pacs{
}

\maketitle



\section{Introduction}
\label{sec:intro}
Weak lensing of the cosmic microwave background (CMB) by the large-scale structure of the Universe converts $E$-mode polarization into $B$-mode, generating an almost white-noise $B$-mode angular power spectrum on large scales equivalent to a noise level of approximately $5\,\mu\text{K}\,\text{arcmin}$. This may confuse a small primordial, gravity-wave induced $B$-mode signal~\cite{Zaldarriaga:1998ar,Hu:2001fb}. At the same time, lensing attenuates the acoustic features of the CMB spectra,  and pushes part of the information encoded in primordial Gaussian fields into higher-order statistics, where it becomes harder to recover. Hence there is clear motivation to delens the CMB we observe. Delensing is still very much in its infancy, but will become increasingly important given the community efforts to detect primordial $B$-mode polarization and hence constrain inflationary gravitational waves~\cite{Hirata:2003ka,Seljak:2003pn,Smith:2008an,Smith:2010gu,Simard:2014aqa}. Delensing the temperature and $E$-mode power spectra can also increase their information content, allowing for better constraints on parameters like the early radiation density~\cite{Green:2016cjr}.

The first demonstration of direct delensing of CMB data was recently published in Ref.~\cite{Larsen:2016wpa}, where the cosmic infrared background (CIB) from star-forming dusty galaxies was used as a tracer of the CMB lensing convergence in order to remap the temperature anisotropies measured by Planck. In this paper, we present a first demonstration of internal delensing, using Planck CMB temperature and polarization measurements themselves to estimate the CMB lensing potential, and then using the estimated deflection map to undo partially the lensing to make delensed maps and power spectra. While the CIB will remain a useful tracer of CMB lensing for some years~\cite{Sherwin:2015baa}, the CIB is difficult to isolate on the largest scales where Galactic dust becomes dominant~\cite{Larsen:2016wpa,Ying:2016eiz}. Once the $B$-mode instrumental noise levels become lower than the lensing-induced power, internal lensing reconstruction also becomes a better tracer of the underlying field, and the ability to delens accurately will become crucial to obtain the most sensitive measurements of the primordial gravitational wave signal.
\newline
\indent
Delensing aims at undoing the lensing in an observed map $T$ (or Stokes parameters $Q$ and $U$ for linear polarization) by remapping points to their undeflected positions. From a deflection estimate $\hat\deflect(\hn)$ this can be done at lowest order by sending $T(\hn)$ to $T(\hn - \hat\deflect(\hn)) \approx T(\hn) - \hat\deflect(\hn) \cdot \nabla T + \cdots$. In contrast to delensing with an external tracer, such as the CIB, when using an internally-estimated lensing map the reconstruction noise in $\hat\deflect$ is no longer independent of the CMB since it is obtained from the same maps. This can lead to significant biases in delensed power spectra, originating from non-zero disconnected correlators like $\la T  \,\hat\deflect \cdot \nabla T \ra $ that would vanish for an external tracer~\cite{Sehgal:2016eag}.
\newline
\indent
To understand the effect of the dependence of the reconstruction noise on the CMB fields, consider reconstructing a large-scale lensing convergence mode using a lensing quadratic estimator. In a local patch, the quadratic lensing estimator works by comparing the locally-measured CMB power spectrum to the full-sky average. If the peaks appear shifted to lower multipoles $\ell$, the patch must be magnified, so the estimate is a positive convergence. If the peaks are shifted to higher $\ell$, the patch must be demagnified, and the estimate is a negative convergence. Now consider a purely Gaussian cosmic variance fluctuation that happens to make the peaks shift to slightly lower $\ell$ in the local patch: the lensing estimator will then return a positive convergence (which is pure reconstruction noise). If we now delens the small patch, using this estimated positive convergence, we will shift the peak back towards the full-sky average. This has the effect of removing random fluctuations in the peak positions, so the full-sky, made of many such delensed patches, will have sharper acoustic peaks than it did before delensing. \revision{Similarly, random fluctuations in the ellipticity of CMB perturbations will be
picked up by the shear part of the reconstruction estimator,
and the delensing operation will undo this `shear' on the map. The perturbations will be more circular after delensing, and therefore the observed spread in scales caused by the random ellipticities will be reduced.} This sharpening of the peaks, which will happen even for lensing-free Gaussian fields, looks very similar to what we would expect from actual delensing, so the naive delensed power spectrum will be strongly biased as the acoustic peaks will be sharpened too much.

Similar biases also appear when internally delensing CMB polarization~\cite{Namikawa:2014yca}. As the signal-to-noise on the reconstruction improves, the biases will also become relatively smaller as the size of relative noise fluctuations goes down, but for current and forthcoming data the biases remain significant.  For Planck, they dominate the delensing signature in magnitude. The biases can be mitigated by filtering in $\ell$ to avoid using non-independent modes~\cite{Teng:2011xc,Sehgal:2016eag}, or they can be modelled. The  biases are difficult to model analytically, non-perturbatively, where they originate from high-order disconnected correlations of the lensed maps.  In this work we use Gaussian, unlensed simulations to estimate and subtract the bias from our measurements on the data, as explained in Sec.~\ref{DataSims}. We show in Sec.~\ref{Results} that this procedure works reliably on simulations, and the spectra of the delensed Planck maps match predictions very well. We also provide an analytical derivation of the leading-order bias in Appendix~\ref{biases},
based on a perturbative expansion in the displacement. These bias predictions are accurate in the case of idealized Planck simulations, and also a good fit to results from the realistic set of simulations provided by the Planck team that we use for our main analysis.
Appendix~\ref{Iterative} presents simulation results showing that using a more
optimal iterative approach to delensing would bring only very minor improvements to the delensing performance at Planck noise levels.
\label{sec:intro}

\section{Methodology \label{DataSims}}
There are several possible implementations of delensing that vary in their details, with their optimality depending upon the precise question being asked. This section introduces the methodology that we adopt. We first discuss our data choices in Sec.~\ref{Data}, and the several suites of simulations that we use are discussed in Sec.~\ref{Sims}. The reconstruction of the lensing potential is discussed in Sec.~\ref{Potential} and our choice of inverse displacement for remapping the CMB fields in Sec.~\ref{Delensing}. Finally, our calculations of the resulting spectra and predictions are given in Sec.~\ref{Spectra}. 

\subsection{Data \label{Data}}

We use the publicly-available foreground-cleaned (SMICA) temperature and polarization maps from the Planck 2015 release\footnote{Maps and masks are available from the legacy archive at \url{http://pla.esac.esa.int/pla/}}~\cite{Adam:2015rua,Adam:2015tpy}. The maps are masked (without apodization) by the SMICA confidence mask, together with the 70\,\% Galactic mask, the point source masks at 143\,GHz and 217\,GHz, and a mask targeted at the resolved Sunyaev-Zel'dovich (SZ) clusters with $S/N>5$ listed in the 2015 SZ catalogue\footnote{\url{https://wiki.cosmos.esa.int/planckpla2015/index.php/Catalogues}} (which has little impact on results). After masking we are left with 67\,\% of the sky.

\subsection{Simulations \label{Sims}}
\newcommand{\setun}{$\rm{S}1$}
\newcommand{\setdeux}{$\rm{S}2$}
\newcommand{\settrois}{$\rm{S}3$}
We use three sets of simulations, labeleled \setun, \setdeux and \settrois. Only \setun\ and \setdeux\ directly enter our baseline analysis and results.

\setun\ consists of 119 Full Focal Plane (labelled FFP9) simulations of the Planck 2015 data, which are publicly available\footnote{\url{https://wiki.cosmos.esa.int/planckpla2015/index.php/Simulation_data}}. These maps form the most realistic set that we use; they contain several layers of Planck-specific systematic effects, e.g., scan strategy, anisotropic beams, anisotropic pixel hit counts etc.\ (see Ref.~\cite{Adam:2015tpy} for details). We use this set of simulations for the lensing potential reconstruction on the SMICA maps.  We also use this set to obtain the covariance matrix of our delensed spectra from the SMICA maps, simply by repeating the analysis on each simulation. The auto-spectra of the FFP9 simulations do not match the data perfectly. In temperature, the mismatch is maximal at our highest multipoles $\ell \approx 1500$ and reaches the $2\,\%$ level, mainly due to residual foreground power after component separation that is not present in the simulations.
The mismatch can reach 5\,\% in $E$-polarization and 5--10\,\% in $B$-polarization, as they are also sensitive to noise modelling errors. Since later on we will be only comparing differences of lensed and delensed spectra, the direct additive component cancels, but the simulations will also slightly misestimate the lensing reconstruction noise leading to a small systematic error in our comparisons of simulations with estimates from data. We found that if unaccounted for this small modelling error would not change our conclusions, with the exception of the $BB$ spectrum result, where good calibration of a modelling bias is required. To account for this power mismatch, additional power has been added to all the simulations used in this paper as isotropic Gaussian noise, using as input spectra that are smooth fits to the measured $TT, EE$ and $BB$ spectral differences between the data and the simulation averages. The way we obtain spectra is described in Sec.~\ref{Spectra}.

\setdeux\ consists of 119 simulations of Planck CMB skies, with the same sky cuts as for the data we are using. The noise simulations are the same (FFP9) as those in \setun,  but not the CMB simulations. The latter are generated from the same theory angular power spectra as for FFP9, but use only effective isotropic transfer functions and no further Planck-specific details. They are not lensed but, rather, are simulated using lensed CMB power spectra. We use this second set of Gaussian simulations to estimate the bias on the delensed spectra, to separate it from true delensing effects.

\settrois\ consists of a similar number of more versatile, faster, flat-sky simulations of idealized Planck-like CMB maps. We use these maps for consistency checks, to test some analysis choices (such as $\ell$ cuts in the data and lensing potential reconstruction) and to explore possible improvements to the analysis.
Some of these tests are described in more detail in Sec.~\ref{Delensing}. For these simulations, the input sky is modelled as a square of area $4\pi$, with power at Fourier wavevector $\vell$ given by the curved-sky $C_{\ell}$ for $\ell$ the integer closest to $|\vell| -1/2$.

In all cases we use the same fiducial $\Lambda$CDM cosmology as the FFP9 simulations, and our fiducial noise spectra are the FFP9 noise spectra including the missing power as described above.

\subsection{Lensing potential reconstruction \label{Potential}}
We reconstruct the lensing potential $\phi$ using just the temperature maps ($\hat \phi^{TT}$), and also the minimum-variance (MV) estimator built from all pairs of maps ($\hat \phi^{\rm{MV}}$).  The quadratic-estimator pipeline estimates the lensing potential from the SMICA maps following very closely the methodology of the Planck 2015 lensing analysis \cite{Ade:2015zua} (hereafter \PlanckLens), to which we refer the reader for full details. The main steps are as follows.
\begin{itemize}
\item Inverse-variance filtering,
\begin{equation}
\label{filter}
\begin{split}
 X^{\rm dat} \rightarrow \left(C^{\rm fid}\right)^{-1}&\left[\left(C^{\rm fid}\right)^{-1} + b^t N^{-1} b\right]^{-1} b^t N^{-1}X^{\rm dat}
\end{split}
\end{equation}
of the masked input maps $X^{\rm \dat} = (T,Q,U)$, using a conjugate-gradient solver with a multigrid preconditioner~\cite{Smith:2007rg} for the large inverse matrix in brackets. This operation down-weights the noisy modes and fills in the masked regions with reconstructed CMB modes. The transfer function ($b$) we use here is a simple isotropic Gaussian beam with full width at half maximum of 5\,arcmin. The noise covariance matrix $N$ is approximated as diagonal in pixel space, with constant noise levels of $N_T = 35\,\mu\text{K}\,\text{arcmin}$ in temperature and $N_P = 55\,\mu\text{K}\,\text{arcmin}$ in polarization for unmasked pixels. The set of fiducial spectra $C^{\rm fid}$ are those of our fiducial cosmology. However, we ignore the $TE$ correlation so that we independently filter the temperature and polarization maps. This approximation results in MV lensing reconstruction noise levels $N_{\ell,0}$ that are suboptimal by an acceptable $2.5\,\%$ at $\ell \sim 100$ to $5\,\%$ at $\ell \sim 2048$.
\item Using these inverse-variance filtered maps in a quadratic estimator to estimate the lensing potential, using fast real-space convolution methods. Only modes $100 \leq \ell \leq 2048$ of these maps are used in the estimator. The lensing maps are then normalized by the analytical lensing response functions.
\item These operations are repeated on all simulations in \setun, and the results averaged, to obtain the mean field $\hat \phi_{\rm{MF}} = \langle\hat \phi\rangle$ , which is subtracted from the potential estimated from the data. This mean-field subtraction suppresses all sources of anisotropies that are not due to lensing and are modelled in the simulations. The mean-field correction is especially important on large scales.
\end{itemize}

\begin{figure}[htp]
\includegraphics[width = 0.5\textwidth]{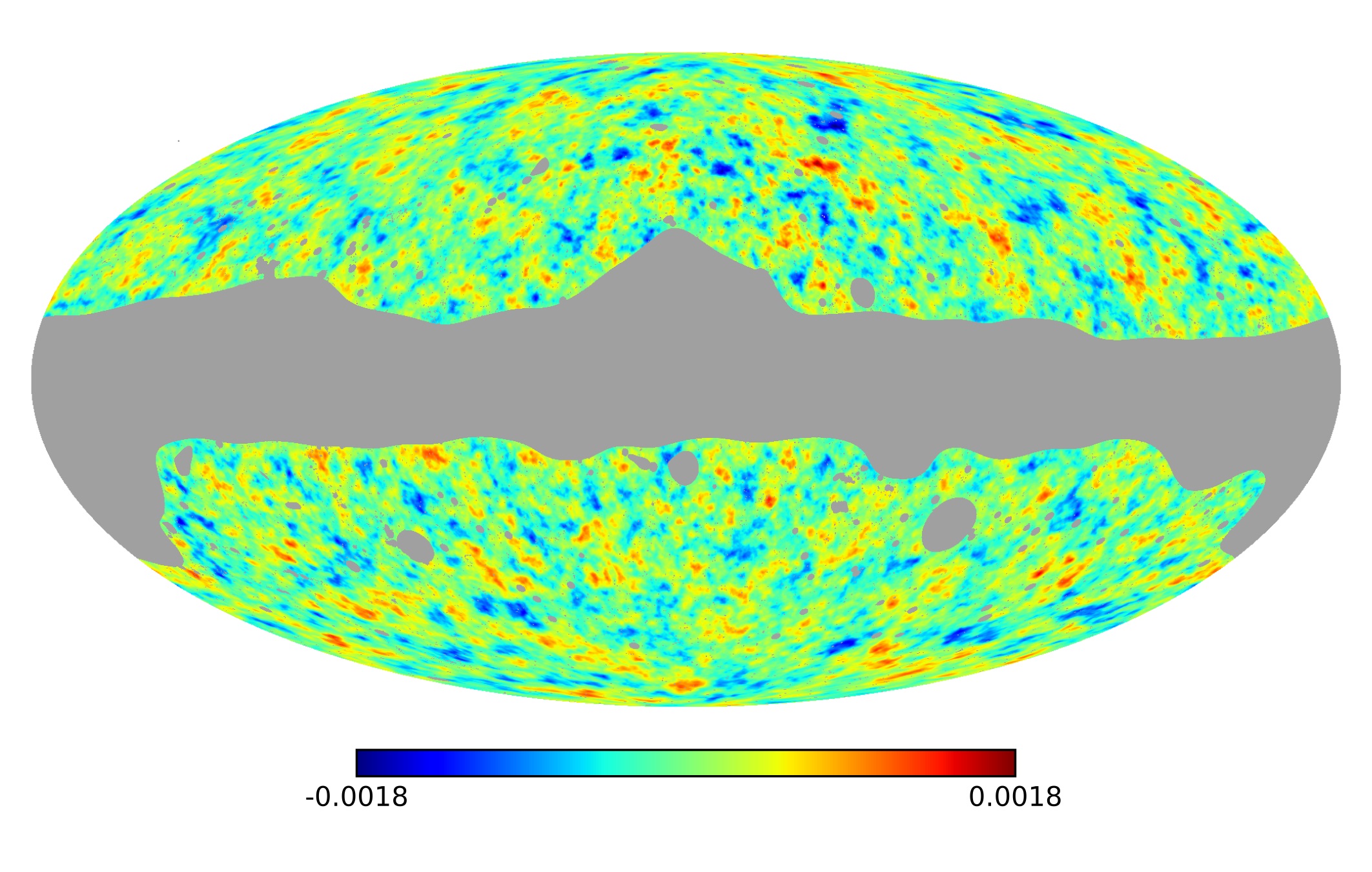}
\caption{ \label{Figdlm} Wiener-filtered minimum-variance displacement field with multipoles $\hat\alpha^{\rm{MV}}_{\clw,\ell m} = \sqrt{\ell(\ell + 1) }\hat \phi^{\rm{MV}}_{\clw,\ell m}$ reconstructed from the SMICA temperature and polarization data maps. We use this reconstruction to delens the Planck $T$, $Q$ and $U$ maps. The mask is shown in grey.
}
\end{figure}

\indent
As a check on our reconstruction of the lensing potential, we build the corresponding lensing potential power spectrum, subtracting the $N_{\ell,0}$ and $N_{\ell,1}$ lensing biases, together with a small Monte-Carlo (MC) correction, following again~\PlanckLens. We find a lensing amplitude with respect to the fiducial model of $0.97 \pm 0.02$ (MV) and $0.98 \pm 0.03$ ($TT$) over the range $8 \leq \ell \leq 2048$, in good agreement with expectations.
\newline
\indent
The quadratic estimate is filtered in multipole space using
\begin{equation}
\label{WF1}
\hat \phi_{\clw,\ell m} = \clw_\ell \:\hat \phi_{\ell  m},
\end{equation}
in order to suppress the noisy small-scale modes. We use the Wiener filter
\begin{equation}
\label{WF2}
\clw_\ell =  \frac{C_\ell^{\rm{fid},\phi\phi}}{C_\ell^{\rm{fid},\phi\phi} +N_{\ell,0} }
 \end{equation}
 throughout. We calculate $N_{\ell,0}$ from the \setun\ simulations. When filtering the reconstruction from the actual data, we use the realization-dependent $N_{\ell,0}$, obtained following~\PlanckLens.
\AC{Still not obvious to me that this is the ``correct'' thing to do.}
\revision{When filtering reconstruction from simulations we use a single Monte-Carlo $N^{\rm MC}_{\ell,0}$ calculated as follows. Splitting the simulation set in two parts, we apply the quadratic estimator to pairs of maps with one map in each set, and average to get the resulting noise spectrum
 \begin{equation}
 N^{\rm MC}_{\ell,0} \equiv 2 \av{C_\ell^{\hat \phi_{12}\hat\phi_{12}}}_{\rm{set}\:1,2}.
 \end{equation}}
 \revision{Usage of the realization dependent bias in the filtering might be useful to reduce the scatter of biases introduced later in Sec.~\ref{Spectra}, but could also in principle introduce undesired complications in the analysis. However, we found that after accounting for the simulation power mismatch as described in Sec.~\ref{Sims}, the realization-dependent $N_{\ell,0}$ agrees with the
Monte-Carlo $N^{\rm MC}_{\ell,0}$ to better than 1\,\%, and therefore the choice of bias in Eq.~\eqref{WF2} when filtering the data has no impact on our results.}
We neglect the $N_1$ lensing bias, which arises from non-primary couplings of the CMB connected 4-point functions to the quadratic estimators, for the purpose of filtering. Figure~\ref{Figdlm} shows the Wiener-filtered displacement field for the MV reconstruction
\begin{equation}
\label{dlm}
\hat\alpha_{\clw,\ell m} = \sqrt{\ell(\ell + 1)} \hat \phi_{\clw,\ell m}.
\end{equation}

\subsection{Delensing}
\label{Delensing}
\indent
Recall that the lensed and unlensed CMB are related by\footnote{On the curved sky, by
``$\hn + \deflect(\hn)$'' we mean parallel transport of distance $|\deflect|$ along the great circle defined by the direction of the local deflection vector $\deflect$~\cite{Challinor02}.}
\begin{equation}
X^{\rm len}(\hn) = X^{\rm unl}(\hn + \deflect(\hn)).
\end{equation}
To delens we want to find the inverse deflection $\invdeflect(\hn)$ defined by
\begin{equation}
X^{\rm unl}(\hn) = X^{\rm len}(\hn + \invdeflect(\hn)).
\end{equation}
With an estimate $\deflect$ of the deflection field in hand, how should we find $\invdeflect$ in order to delens?

The most obvious answer, but not necessarily the most practical, is to solve explicitly for the inverse deflection. The deflection fields (and especially the Wiener-filtered ones) are weak enough in CMB lensing that the lensing-induced remapping of the points on the sky is one-to-one. Formally, this is always the case as long as the magnification matrix
\begin{equation}
\label{magn}
M_{ab}(\hn) = g_{ab} + \nabla_a \alpha_b(\hn)
\end{equation}
is invertible. (Here, $g_{ab}$ is the metric on the sphere and $\nabla_a$ is the covariant derivative.) This requires (perturbatively) $|2 \kappa | \ll 1$ at each point, where the lensing convergence $\kappa = -\nabla_a \alpha^a/2$,
which is easily satisfied at our resolution if the reconstruction noise has been filtered. A more practical solution is to use a simple approximation for the inverse, such as $ \invdeflect\approx-\deflect$ (dubbed `anti-lensing' by Ref.~\cite{Anderes:2014foa}), with errors going like $\deflect \cdot \nabla \deflect$ at leading order. We check the impact of this approximation using the faster flat-sky set of simulations \settrois.

The requirement that each point $\hn$ is remapped onto itself provides an equation for the inverse deflection:
\begin{equation}
\hn + \invdeflect(\hn) + \deflect(\hn + \invdeflect(\hn)) = \hn.
\end{equation}
Note that even if the forward displacement is a pure gradient, the inverse displacement will have a small curl component and has two genuine degrees of freedom. We obtain the inverse iteratively, using a simple Newton-Raphson scheme. Operating at resolution 0.7\,arcmin, we iterate
\begin{multline}
\invdeflect_{N + 1}(\hn) = \invdeflect_N(\hn) \\
- \boldvec{M}^{-1}(\hn + \invdeflect_N(\hn)) \left[ \invdeflect_N(\hn) + \deflect(\hn + \invdeflect_N(\hn))\right]
\end{multline}
starting from $\invdeflect_0 \equiv 0$, where $\boldvec{M}$ is the magnification matrix defined in Eq.~\eqref{magn}. The necessary interpolations are performed with standard bicubic spline techniques. Three iterations are typically sufficient for satisfactory convergence.
\newline
\indent
Building error histograms, the accuracy of the approximation $\invdeflect(\hn) = -\deflect(\hn)$ for a noise-free $\Lambda $CDM displacement has a root-mean-square (RMS) error of 17\,\%, with errors comparable to unity in the tails. On the other hand, in the more relevant case of the displacement filtered using Eqs. \eqref{WF1} and \eqref{WF2}, using Planck  $N_0$, the RMS drops to $1.3\,\%$ ($\hat\phi^{TT}$), or $1.9\,\%$ ($\hat \phi^{\rm{MV}}$), and errors are nowhere larger than 10\,\% (15\,\% for MV). We see no significant differences at the level of the delensed spectra. Hence, we use $-\deflect$ as the inverse displacement for the main results of this paper, using our filtered deflection field to estimate the delensed field as
\begin{equation}
X^{\rm del} \equiv X^{\rm dat}(\hn - \hat\deflect_{\clw}(\hn)).
\end{equation}
We use the full range $1 \leq \ell \leq 2048$ of the potential reconstruction for this purpose. The potentially poorly-understood low-ell modes do not affect significantly our results, as discussed in more details in Sec.~\ref{Results}.
\subsection{Spectra and predictions \label{Spectra}}
\indent
The maps we delens are built as follows. We start with the CMB maps filtered using Eq.~\eqref{filter} that we also use for the lensing reconstruction, with multipole cuts $100 \leq \ell \leq 2048$. We then rescale these maps at each multipole by the relevant isotropic limit of the inverse filtering. Since we neglect $C_\ell^{TE}$ in the filtering and use constant pixel noise, this is simply $(C^{TT,\rm fid} + N_T/b_l^2)$ in temperature and $(C^{EE,\rm fid} + N_P/b_l^2)$ and $(C^{BB,\rm fid} + N_P/b_l^2)$ in polarization. Away from the mask these maps are simply the beam-deconvolved, band-pass filtered data. Due to the extent of the mode-coupling due to lensing (typically a few hundred multipoles), we further discard multipoles $\ell > 1500$. This, however, conserves almost the entire signal-to-noise of our results. Aside from these sharp $\ell$-cuts we apply no additional filtering.
\newline
\indent
The maps are delensed by remapping points, in the same way that lensed maps are simulated: using  a degree-7 bivariate barycentric Lagrange interpolation \cite{BarycentricLagrange} on the equatorial cylindrical projection of the map. We obtain estimates for the temperature and polarization power spectra before ($\hat C^{\dat}_\ell$) and after ($\hat C^{\del}_\ell$) the delensing operation. Since the masked pixels were filled in by the filtering we do not apodize or further deconvolve the mask, but simply use the naive power spectrum estimate
\begin{equation}
\hat C_\ell = \frac{1}{(2\ell + 1)f_{\rm sky}}\sum_{m} |a_{lm}|^2 ,
\end{equation}
where $f_{\rm{sky}} \approx 0.67$.
We then build the combination $\hat C_\ell^{\del} - \hat C^{\dat}_\ell$ to quantify the delensing effect, which cancels out the bulk of the instrument noise and CMB cosmic variance, especially at lower $\ell$ where the lensed and delensed fields remain highly correlated. The naive $1/ f_{\rm{sky}}$ recipe to obtain the spectra is inaccurate in detail, with biases amounting to several percent on the scales relevant here.
However, since we are differencing spectra calculated in the same way, this only gives a percent-level error on the difference; this is modelled consistently in our simulations, and is acceptably small compared to our error bars.
\revision{
The filtering step
 fills the mask with reconstructed CMB modes, but away from the mask edges these are only large scale and do not contribute significantly to the lensing reconstruction after mean field subtraction. Although the mask is unapodized, the transition across the mask edge is smooth for the filtered map, so the mask does not significantly complicate the delensing procedure and any edge effects are self-consistently modelled in our simulations.
Furthemore, we show in Sec.~\ref{Results} that our results are in very good agreement with expectations from idealized simulations, so we can be confident that are results our robust to the impact of sky cuts.
}

\begin{figure}[htp]
\includegraphics[width = 0.49\textwidth]{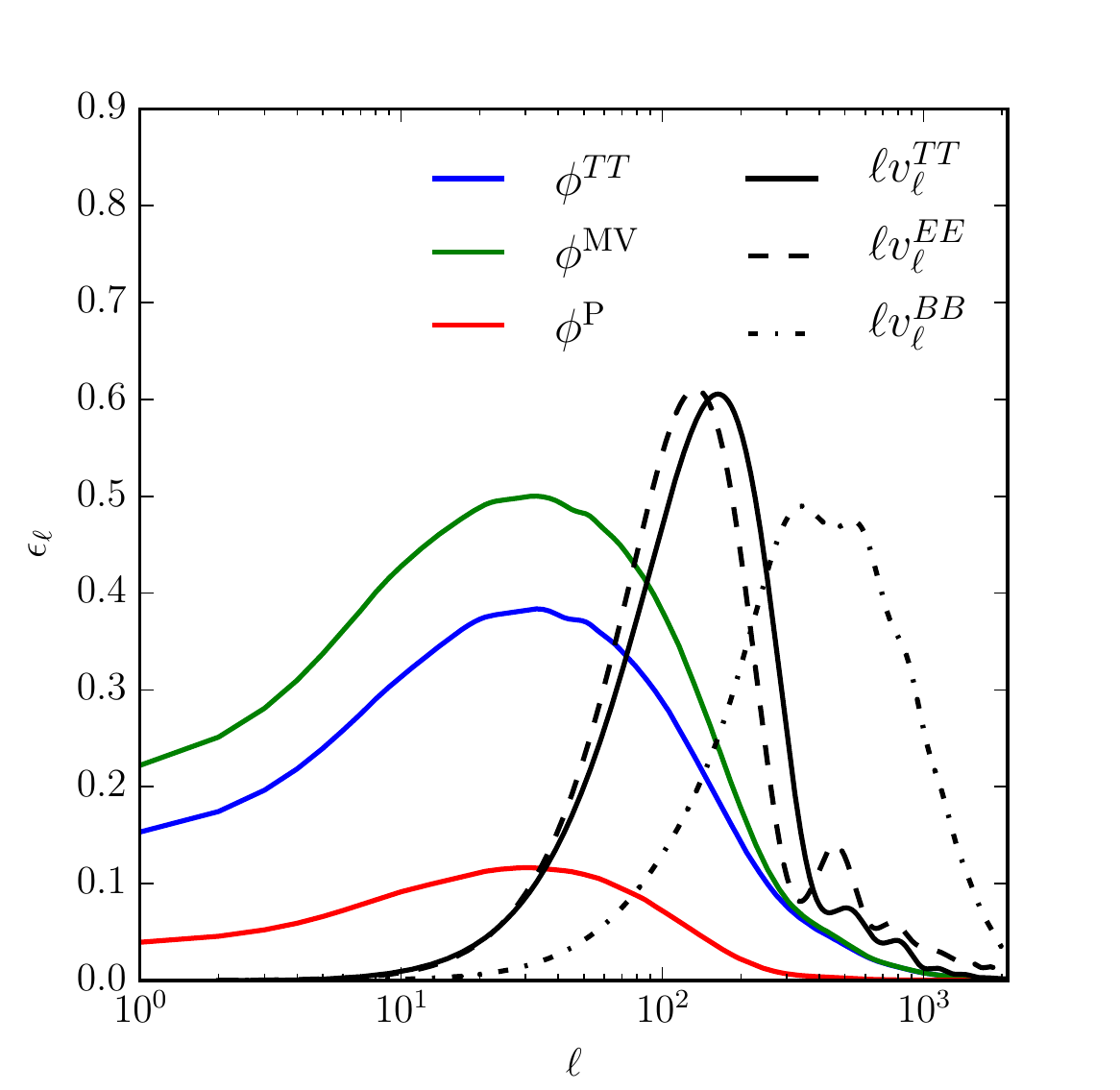}
 \caption{ \label{Figeps}
Expected delensing efficiencies $\epsilon_\ell = 1- C^{\phi\phi}_{\ell,\rm{del}} / C^{\phi\phi}_\ell$
[see Eq.~\eqref{resLP}]
achievable given SMICA noise levels for $TT$ (blue) and MV (green) lens reconstructions.
Also shown for comparison is the case of a polarization-only reconstruction ($\phi^{\rm{P}}$; red) combining the $EB$ and $EE$ quadratic estimators. The black lines show the contribution per log $\ell$ of the deflection power to the lensed CMB power spectra, approximated as the leading right eigenvector $v^{XY}_\ell$ of the singular value decomposition of $\ud C_\ell^{XY}/\ud(\ln  C_L^{\phi\phi})$ (evaluated using the {\sc lenscov} package~\cite{Peloton:2016kbw} for $\ell_{\rm max}=2448$).
The result for $TE$ is very close to that for $EE$ and is not shown separately. The total CMB power spectrum delensing efficiencies are given approximately by an average of the efficiency curves in color weighted by the relative size of the black curves.}
\end{figure}

The estimate of the lensing field that we use to delens is both filtered and noisy, so schematically we have
\begin{multline}
X^{\rm del} = X^{\rm dat}(\hn - \hat\deflect_\clw)
= X^{\rm dat}(\hn - \clw \star\left ( \deflect+ \boldsymbol{n}_0\right))\\
 \approx X^{\rm unl}\left(\hn + (1-\clw)\star\deflect - \clw\star\boldsymbol{n}_0\right) + \text{noise},
\end{multline}
where $\boldsymbol{n}_0$ is the realization of the reconstruction noise, ``$\text{noise}$'' is the delensed instrumental noise, and $\star$ denotes convolution.
The delensed field is therefore equivalent to having the unlensed CMB lensed by residual deflections
\begin{equation}
\deflect - \hat\deflect_\clw = (1-\clw) \star \deflect - \clw \star \boldsymbol{n}_0,
\label{eq:residual}
\end{equation}
which have a power spectrum determined by
\begin{equation}
\label{CAMB}
C_{\ell,\rm{del}}^{\phi\phi} =
\left ( 1 - \clw_\ell \right)^2 C_\ell^{\phi\phi} +  \clw^2_\ell N_{\ell,0}.
\end{equation}
The $\clw_\ell$ that minimizes this residual power is the Wiener filter of Eq.~\eqref{WF2}, and the result then simplifies to
\begin{equation}
\label{resLP}
C_{\ell,\rm{del}}^{\phi\phi}=
\left ( 1 - \clw_\ell \right) C_\ell^{\phi\phi}.
\end{equation}
\newline
\indent
Hence $\clw_\ell$ is the maximal, multipole-dependent delensing efficiency we can expect to be able to achieve with the quadratic estimator. For a generic tracer $\hat \phi$, with arbitrary auto and cross-spectrum with $\phi$, the maximal efficiency generalizes to
\begin{equation}
\label{CorrEps}
\epsilon_\ell \equiv  1- C^{\phi\phi}_{\ell,\rm{del}} / C^{\phi\phi}_\ell = \frac{\left(C^{\hat \phi \phi}_\ell\right)^2}{C^{\phi\phi}_\ell C^{\hat \phi \hat \phi}_\ell}.
\end{equation}
The correlation coefficient $\rho_\ell$ between the tracer $\hat{\phi}$ and the true lensing signal (as used in Ref.~\cite{Larsen:2016wpa}) is simply related by $\epsilon_\ell = \rho_\ell^2$.
The $\epsilon_\ell$ are shown in Fig.~\ref{Figeps} for the $TT$, ${\rm MV}$ and a polarization-only reconstruction with Planck data.

If the reconstruction noise were independent of the unlensed CMB, $C_{\ell,\rm{del}}^{\phi\phi}$ would directly determine the expected delensed CMB power spectrum signal that we expect to see (via the usual calculation of the lensed spectra~\cite{Lewis:2006fu} in terms of the power spectrum of the deflection angles that are independent of the unlensed CMB).
However, using internal delensing, the reconstruction noise is not independent of the CMB fields being delensed. So, although Eq.~\eqref{resLP} correctly quantifies the residual deflection, it does not directly determine the delensed spectra: compared to the naive result we see significant power spectrum biases arising from the non-independence of the CMB and reconstruction noise.

When considering delensed power spectra it is useful to consider the debiased estimator
\begin{equation}
\label{Combi}
\debiasedChat  \equiv
\hat C^{\del}_\ell - B_\ell,
\end{equation}
where the `bias' $B_\ell$ is a disconnected (Gaussian) bias that arises even when there is no lensing. We can define it by applying the same series of operations on Gaussian maps that are not lensed, but have the same spectra as the lensed CMB:
\begin{equation}
\label{Bdef}
B_\ell \equiv \left.\av{ \hat C^{\del}_\ell - \hat C_\ell^{\dat}  }\right|_{\textrm{Gaussian CMB}}.
\end{equation}
The debiased $\hat C_{\ell,\rm{debias}}^{\rm del} $ therefore has expectation value equal to that of the original $C_\ell^{\dat}$ if there is no non-Gaussian lensing signal.

We can evaluate $B_\ell$ in a fiducial model in order to subtract it.\footnote{
A potentially better realization-dependent estimator could be constructed using the observed lensed power spectrum $\hat{C}_{\ell}$ using
$\hat{C}_{\ell,\rm{debias}}^{\rm{del,RD}} = \hat{C}_\ell^{\rm{del}} - B_\ell - \sum_{\ell '}\frac{\partial B_\ell}{\partial C_{\ell'}}(\hat{C}_{\ell'}-C^{\rm fid}_{\ell'})$ (or equivalent anisotropic generalization), which would also be insensitive to leading-order deviations from the assumed fiducial model.}
We give an analytic calculation in Appendix~\ref{biases} in the idealized full-sky case, and show that in simple cases (like delensing the $TT$ spectrum using the temperature lensing reconstruction) the bias amounts to an additional delensing-like peak sharpening effect that has nothing to do with the actual lensing in the map. The combination of Eq.~\eqref{Combi} also cancels (in the assumed fiducial model) the mean residual lensing smoothing generated by the quadratic estimator noise.
Because of this, the mean residual lensing effect on the CMB signal debiased power spectra is determined by the usual formula with the lensing power reduced from Eq.~\eqref{resLP} to just the first term in Eq.~\eqref{CAMB}:
\begin{equation}
C_{\ell,\rm{del, debias}}^{\phi\phi}=
\left ( 1 - \clw_\ell \right)^2 C_\ell^{\phi\phi}.
\label{eq:CAMBsignalonly}
\end{equation}
This should not be confused with the delensing efficiency of Eq.~\eqref{resLP}, which determines the map-level efficiency of the delensing.

The debiased spectrum of Eq.~\eqref{Combi} also cancels the power due to delensing of the instrumental noise by the quadratic estimator reconstruction noise. However, a component from delensing of the noise by the true sky lensing signal does remain. This signature is relevant for our $BB$ polarization results. Just as for the bias $B_\ell $, it can be evaluated in the fiducial model analytically, or, more realistically, with simulations. In analogy with the definition in Eq.~\eqref{Bdef}, we define a noise bias using simulated noise maps (with the same cuts as the data) delensed by a Wiener-filtered Gaussian lensing potential with spectrum $C_\ell^{\phi\phi,\rm fid}$:
\newcommand{\BnoiseOnly}{B^{\rm{noise}}_{\phi_{\clw},\ell}}
\newcommand{\BcmbOnly}{B^{\textrm{CMB}}_{\hat \phi_{\clw},\ell}}
\begin{equation}
\label{Bnoise}
\BnoiseOnly \equiv \left.\av{ \hat C^{\del, \rm{noise}}_\ell - \hat C_\ell^{\dat, \rm {noise} }}\right|_{\clw \star\phi}.
\end{equation}
\newline
\indent
Finally, we comment on more optimal methods to extract the lensing potential that could increase the delensing efficiency $\epsilon_\ell$. It has long been suggested~\cite{KnoxSong2002,Kesden:2002ku}, and also explicitly demonstrated \cite{HirataSeljak2003a}, that higher-order statistics of the CMB may help the measurement of the lensing potential and improve on the quadratic estimator. This is because the quadratic estimator reconstruction noise levels are set by the lensed spectra of the data, while any fundamental limit is expected to be determined by the cosmic variance of the unlensed spectra \cite{Hirata:2003ka}. However, substantial improvements are not expected for the Planck data, where the polarization is noise dominated. Quantitative details together with a rigorous check of this assertion are provided in Appendix~\ref{Iterative}.

\begin{table*}
\caption{\label{tab:table1}  Detection significances of the reduction of true lensing effects of $\phi$ in the temperature and polarization power spectra, as measured by the amplitudes $A$ [see Eq.~(\ref{eq:significance})], and delensing efficiencies $\epsilon$ [see Eq.~(\ref{template})]. The latter measures the efficiency at which delensing reduces the power spectrum of the net deflections, including reconstruction noise, that remain in the maps after delensing. \revision{For instance, delensing with a reconstruction noise spectrum equal to $C^{\phi\phi}_\ell$ would result in no net delensing and a delensing efficiency of zero.} Results are given for the $TT$ and MV lens reconstructions. All results are based on comparison with lensed simulations, except for the ``null hypothesis'' significances, which instead use the statistics of the amplitude estimator across unlensed simulations (but using the expected template shape from lensed simulations). For the delensing efficiencies, we show the predictions from simulations and from direct calculation with \CAMB\ (see text). We also quote reduced chi-squared values, $\chi^2_r$, based on Eq.~(\ref{template}) -- but using binned spectra -- at the best-fitting efficiencies; the expectations based on Gaussian statistics for the binned spectra are $\left\langle {\chi_r^2}\right\rangle = 1 \pm 0.24$ ($TT,TE,EE$) and $1 \pm 0.33$ ($BB$) We also give the zero-point efficiencies, $\epsilon_0$, expected in the absence of lensing, arising from independent reconstruction noise. Results are quoted to slightly more precision than justified by their Monte-Carlo and known systematic error to allow easier comparison of relative changes in the size of the error bars.}
\resizebox{\textwidth}{!}{%
\begin{tabular}{ lllll  }
\hline
\hline
& $C_\ell^{TT}$& $C_\ell^{TE}$& $C_\ell^{EE}$& $C_\ell^{BB}$\\
\hline
$\hat \phi^{TT}$-delensing  amplitude $A$ & $1.083\pm 0.058$  & $1.016\pm 0.056$ & $0.961\pm 0.084$  & $0.689\pm 0.273$ \\
Significance &$18.7\,\sigma$ & $18.0\,\sigma$ & $11.5\,\sigma$& $2.5\,\sigma$ \\
Significance (null hypothesis)& $22.3\,\sigma$ & $21.4\,\sigma$ & $14.5\,\sigma$ & $3.1\,\sigma$ \\
\hline
$\hat \phi^{\rm{MV}}$-delensing amplitude $A$  & $1.097\pm 0.054$ & $0.987\pm 0.054$  & $0.908\pm 0.079$   & $0.984\pm 0.258$  \\
Significance & $20.2\,\sigma$ &$18.2\,\sigma$ &$11.5\,\sigma$ &$3.8\,\sigma$ \\
Significance (null hypothesis)& $25.6\,\sigma$ & $24.0\,\sigma$ & $15.7\,\sigma$ & $4.6\,\sigma$ \\

\hline
$\hat \phi^{TT}$-delensing efficiency $\epsilon$   & $0.222\pm 0.020$ ($\chi^2 = 1.14$)  & $0.216\pm 0.020$ ($\chi^2 = 0.89$)  & $0.187\pm 0.030$ ($\chi^2 = 1.18$)  & $0.016\pm 0.042$ ($\chi^2 = 0.85$) \\
Prediction (FFP9 simulations) & 0.191& 0.209   & 0.206&0.083  \\
Prediction, leading order (\CAMB)  & 0.236  & 0.237  & 0.245  &0.097\\
Zero-point efficiency $\epsilon_0$   & -0.154   & -0.152   & -0.149  &-0.048  \\


\hline
$\hat \phi^{\rm MV}$-delensing efficiency $\epsilon$& $0.289\pm 0.023$ ($\chi^2 = 1.25$)  & $0.254\pm 0.024$ ($\chi^2 = 0.83$)  & $0.226\pm 0.035$ ($\chi^2 = 0.95$)  & $0.071\pm 0.048$ ($\chi^2 = 0.44$)  \\
Prediction (FFP9 simulations) & 0.246 & 0.260   & 0.270  & 0.1\\
Prediction, leading order (\CAMB)  & 0.303  & 0.303  & 0.314  &0.118\\
Zero-point efficiency $\epsilon_0$   & -0.175   & -0.173   & -0.167   & -0.037 \\



\hline
\end{tabular}
}
\end{table*}

\begin{figure*}[t]
\includegraphics[width = \textwidth]{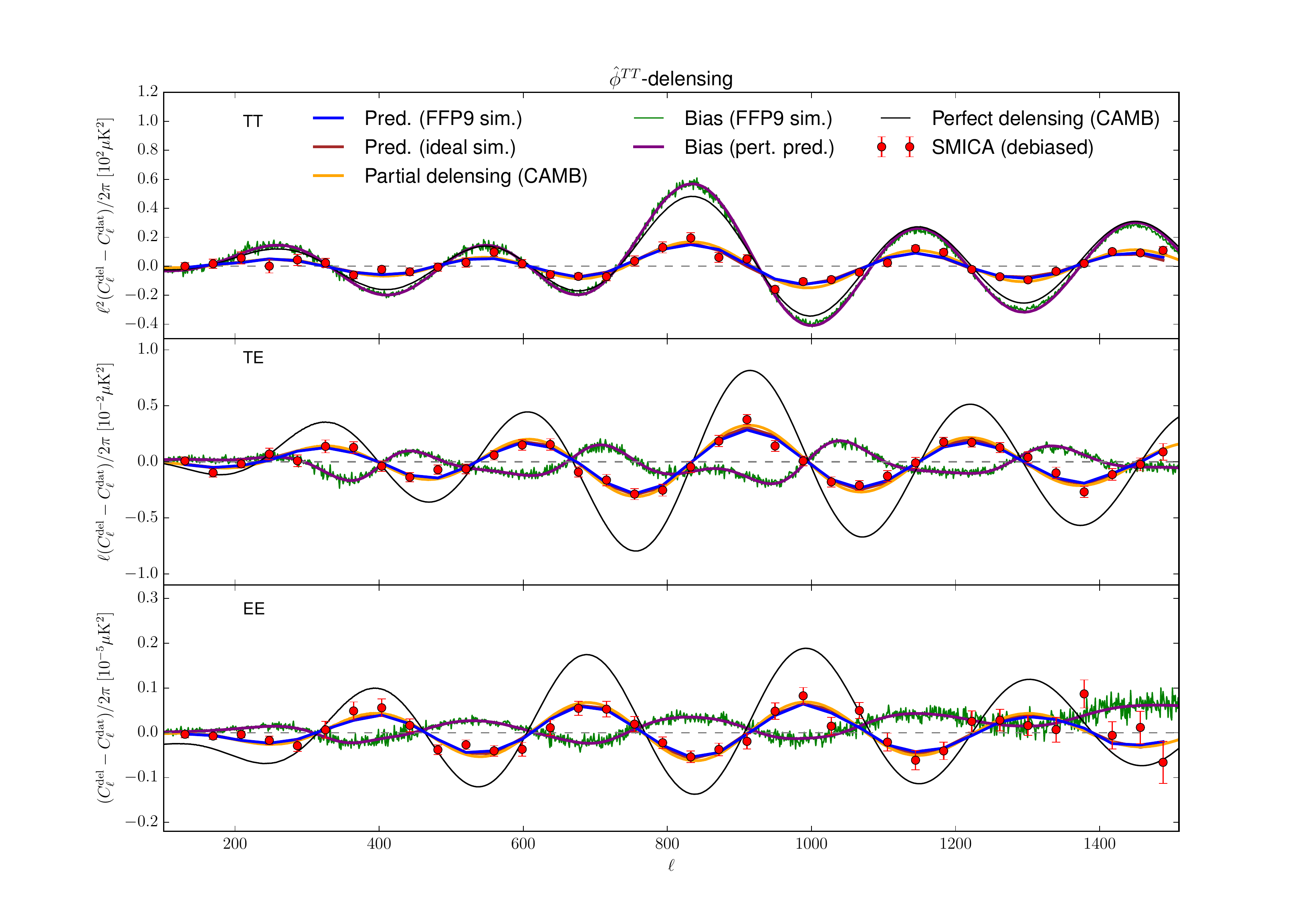}

\caption{Difference $\hat C^{\del}_{\ell,\rm{debias}} - \hat C^{\dat}_\ell - \BnoiseOnly$ of the power spectra of the Planck SMICA maps after delensing to their values before delensing (points and error bars), for $C_\ell^{TT}$, $C_\ell^{TE}$ and $C_\ell^{EE}$ from top to bottom. The maps are delensed using the Wiener-filtered temperature-based lensing potential reconstruction (see Fig.~\ref{Figp} for the case of the minimum-variance lensing potential). The delensed spectra are corrected for the Gaussian bias $B_\ell$ estimated from simulations (green). They are also corrected for the noise delensing $\BnoiseOnly$. The noise delensing bias is essential only for $B$-polarization and is not shown separately here (it is shown in Fig.~\ref{figBiasesPred} of Appendix~\ref{biases}). \revision{It vanishes in the null hypothesis of no lensing signal in the maps.} \revision{Hence, under the null hypothesis, the points on this figure would all be consistent with zero to a very good approximation, and the quoted significance of our results can be read directly from this figure. However, the delensing efficiencies given in Table~\ref{tab:table1} are smaller than this figure suggests, since the peak smoothing by the reconstruction noise has been subtracted by the debiasing procedure.} The Gaussian bias is well modelled by a simple low-order perturbative expansion applied to an isotropic survey (see Appendix~\ref{biases}), as shown in purple. The predictions for the expected spectral difference obtained from the FFP9 Planck simulations are shown in blue. The expected spectral difference computed from a set of idealized, isotropic full-sky Planck simulations with matching power are shown in brown, and agree very well for these spectra with the predictions from the FPP9 simulations. Also shown are leading-order predictions from \CAMB\ (orange) based on Eq.~(\ref{eq:CAMBsignalonly}). The \CAMB\ prediction for complete delensing  $C_\ell^{\text{fid},\rm{unl}} - C_\ell^{\text{fid},\rm{len}}$ is shown in black.
See Table \ref{tab:table1} for fits and $\chi^2$.
\label{Figptt}}
\end{figure*}

\begin{figure*}[t]
\begin{center}
\includegraphics[width = \textwidth]{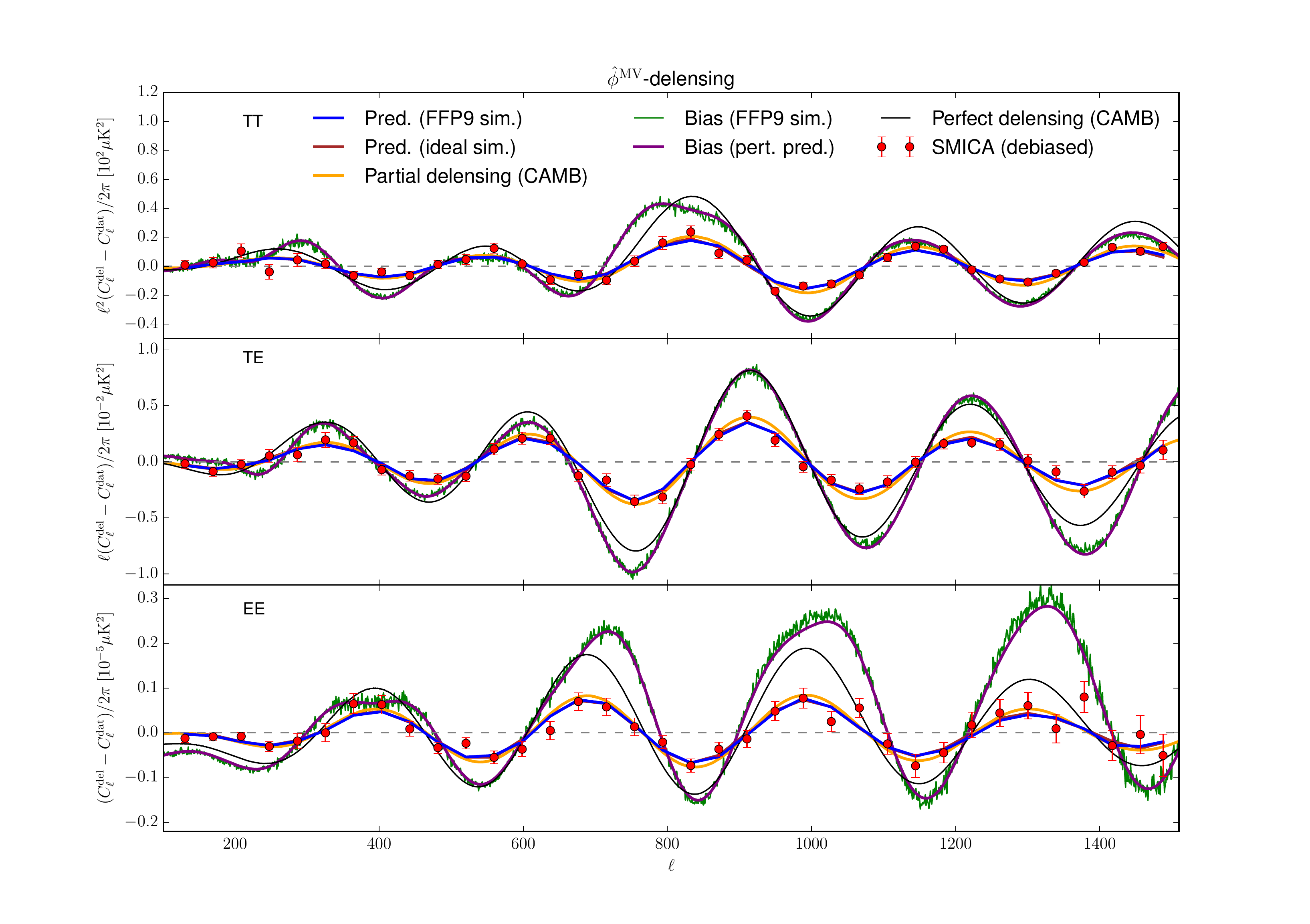}
\end{center}
 \caption{As Fig. \ref{Figptt}, but using the Wiener-filtered minimum-variance estimator $\hat \phi^{\rm{MV}}$ to delens the maps. \label{Figp}}
\end{figure*}

\begin{figure}[t]
\includegraphics[width = 0.98\columnwidth]{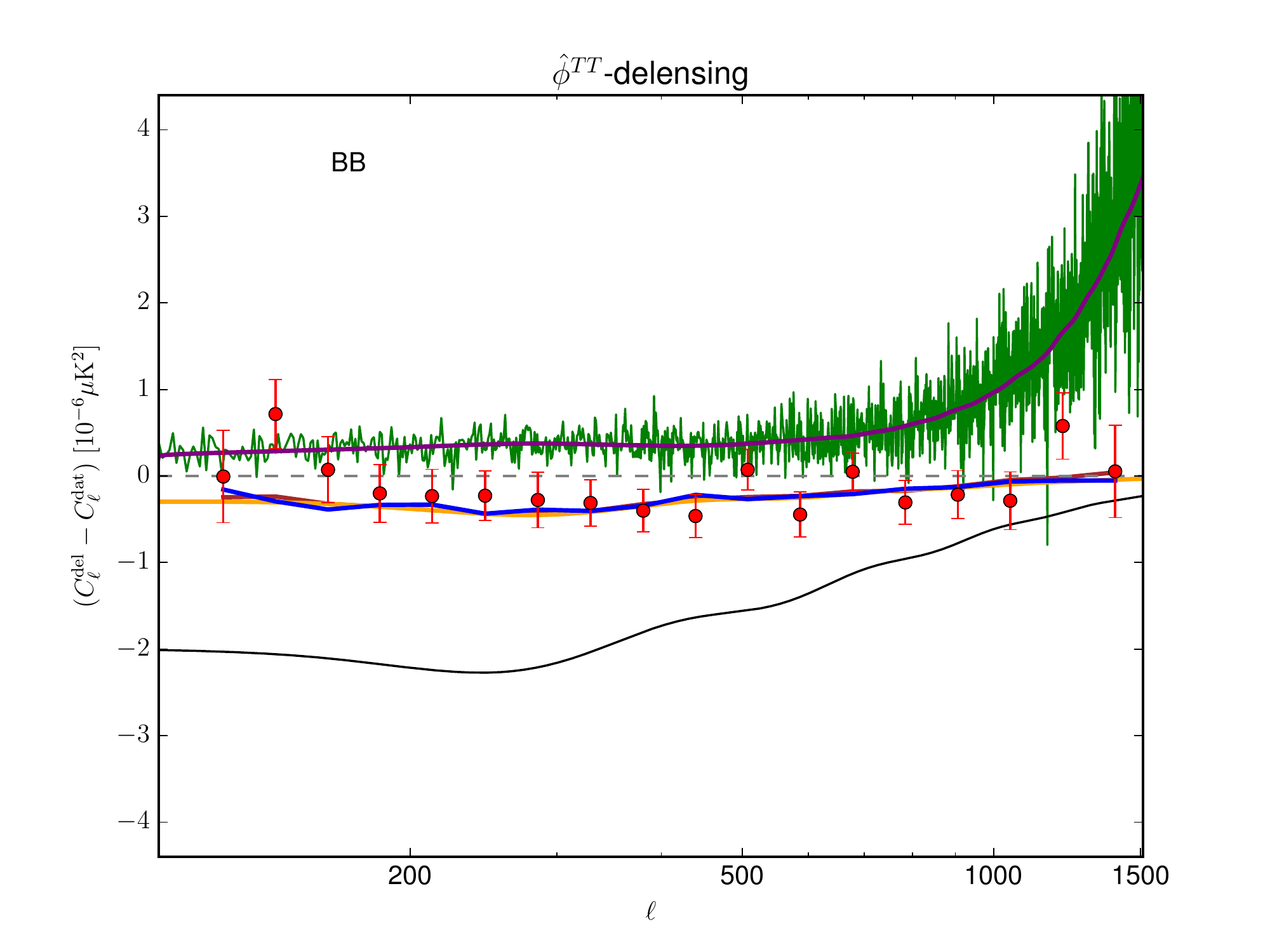}
\includegraphics[width = 0.98\columnwidth]{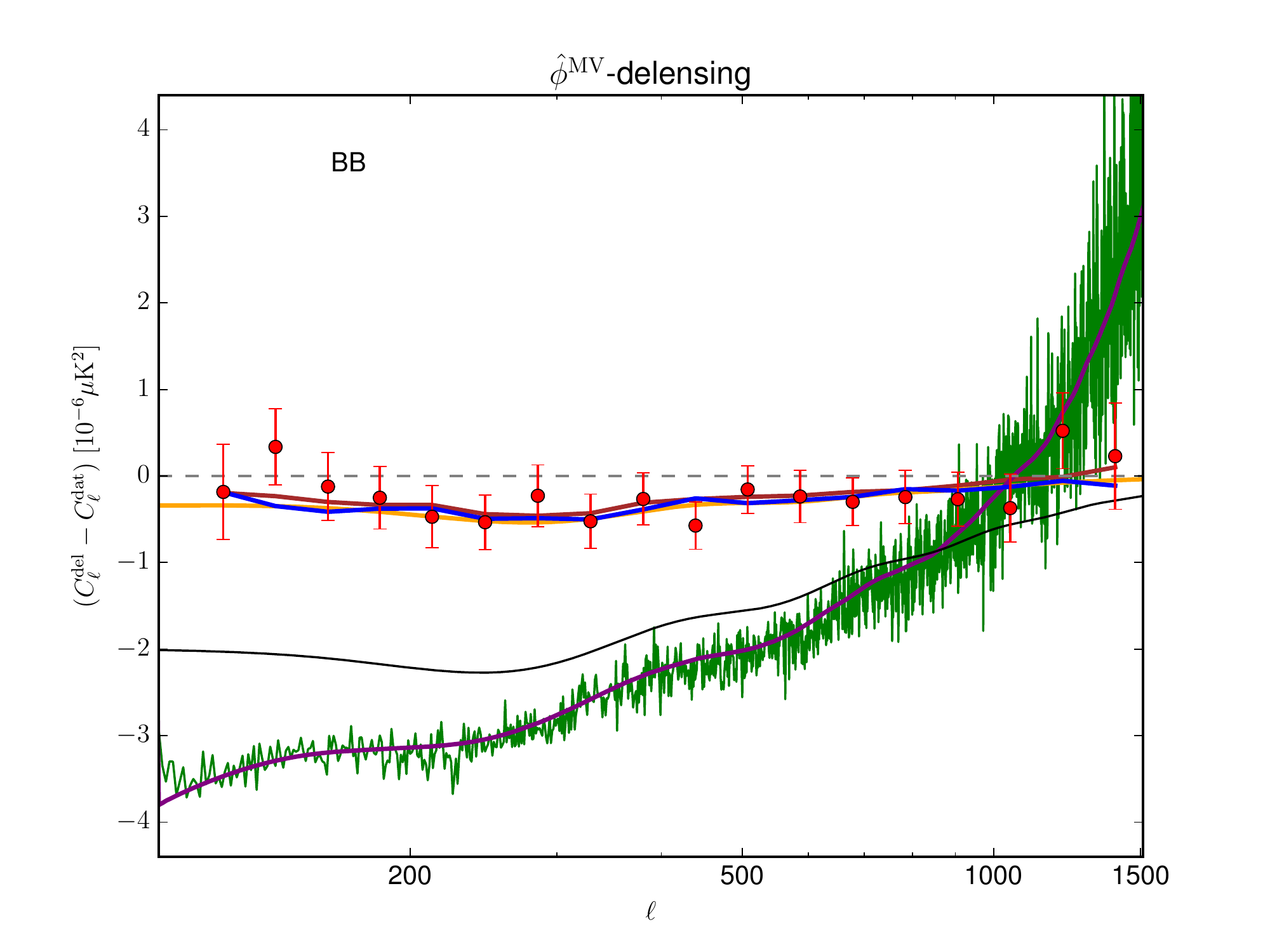}

\caption{As Fig.~\ref{Figptt}, but for the delensed $BB$ spectrum. Results are shown for delensing with the $TT$ (top) and MV (bottom) reconstructions.
Note the logarithmic scale for the multipole axis. \revision{In constrast to Fig.~\ref{Figptt}, the noise delensing bias $\BnoiseOnly$ is important at high multipoles; see Fig.~\ref{figBiasesPred} of Appendix~\ref{biases}.}
\label{FigBB}}
\end{figure}

\section{Results \label{Results}}
Our main results are shown in Fig.~\ref{Figptt} (the impact of delensing with $\hat \phi^{TT}$ on the $TT$, $TE$ and $EE$ spectra), Fig.~\ref{Figp} (the same for delensing with $\hat \phi^{\rm{MV}}$) and Fig.~\ref{FigBB} (the $BB$ spectra with $TT$ and MV delensing). These figures show the result of the delensing procedure applied to the Planck SMICA maps as the data points, using the noise-cancelling combination
$\debiasedChat - \hat C_\ell^{\dat} -\BnoiseOnly$.
The spectra are corrected for the Gaussian bias $B_\ell$, shown in green in the figures, estimated from the Gaussian subset of S2 simulations  (see Sec.~\ref{Spectra}). \revision{By construction, the measured spectra $\debiasedChat - \hat C_\ell^{\dat}$ would be consistent with zero in the absence of lensing in the data maps.} For easier visual comparison to analytical expectations
in $B$ polarization, the spectra are also corrected for the noise delensing bias $\BnoiseOnly$, estimated from the subset \setun\ (FFP9) of noise simulations applied to Eq.~\eqref{Bnoise}. \revision{This subtraction only has a small visual effect on the spectra built from $T$ and $E$ shown on Fig.~\ref{Figptt}}.
\AL{this doesn't really cover AC's point of principle, e.g. in caption of Fig 3 where we say "points would all be consistent with zero", it's not strictly true? (unless you specify in the null hypothesis where Bnoise is zero)}
The black lines show the result for perfect delensing, $C_\ell^{\rm{fid,unl}} - C_\ell^{\rm{fid,len}}$, while the predicted amount of delensing, after bias correction, is shown in blue. These predictions are obtained by delensing the FFP9 S1 simulations, including the same sky cuts and multipole cuts.
We also show predictions obtained by delensing the ideal, full-sky and isotropic Planck simulations S3, as the brown lines. They are generally in very good agreement, demonstrating nicely that the non-ideal effects such as sky-cuts, anisotropic noise and beams, or mean-field contamination have a small impact on the analysis. Finally the orange lines in the figures are the purely analytic expectations discussed in Sec.~\ref{Spectra}, showing the difference from CMB spectra lensed by $(1 - \clw_\ell)^2C_\ell^{\phi\phi}$ (computed with \CAMB). Their deviations to the brown curves show where our simple additive debiasing scheme in Eq.~\eqref{Combi} does not recover accurately the delensed spectrum. As the figures show, this is nowhere a large effect. A more detailed discussion is given in Appendix \ref{biases}, but we do not quantitatively investigate higher-order contributions in this paper.
\newline
\indent
The importance of the bias is striking. The magnitude of the bias is typically larger than the signal itself, and is mostly due to the non-independence of the reconstruction noise and the CMB maps. This dependence gives additional terms in the delensed power spectrum that depend on the disconnected 4-point (and higher-point) moments of the lensed CMB.
In the case of delensing the temperature with $\hat{\phi}^{TT}$, the bias leads to additional peak sharpening (as explained qualitatively in Sec.~\ref{sec:intro} and as an analytic limit in Appendix~\ref{biases}). This is also seen in polarization delensing using the MV reconstruction (Fig.~\ref{Figp}; where the reconstruction depends both on temperature and polarization). The large noisy polarization-polarization peak sharpening bias dominates the smaller temperature-polarization lensing bias because the correlation between the fields is fairly weak.
\newline
\indent
Note that even when the reconstruction noise and unlensed CMB are independent there is still a somewhat smaller contribution to the bias that arises from disconnected six-point (and higher-point) correlators of the lensed CMB. Physically, an independent noise on the deflection angles effectively appears in the delensed maps as an additional residual lensing effect, as though they were lensed by the noise deflection field. In particular, the bias visible in the $BB$ spectrum (see Fig.~\ref{FigBB})
when delensed with $\hat\phi^{TT}$ is expected to come exclusively from this type of correlations, since the $TB$ correlations are zero. This term is also the dominant contribution to the $\hat\phi^{TT}$ delensed $EE$ spectrum.
\newline
\indent
We show in Appendix~\ref{biases} that the analytic perturbative prediction for the bias is very accurate when compared to idealized \settrois\ simulations.  To build analytic predictions for the realistic case, we use input noise spectra based on a smooth fit to a few hundred FFP9 $T$, $Q$ and $U$ noise simulation spectra, including the spectral mismatch correction discussed in Sec.~\ref{Data}.
The corresponding analytic bias predictions are shown as the purple lines in Figs.~\ref{Figptt}--\ref{FigBB}, and provide a good fit the simulation results (shown in green). Some small differences with the simulation results remain, so for better accuracy we use the biases measured from the \setdeux\ simulations for our main results derived from the Planck data.
\newline
\indent
To assess the significance with which we detect the expected changes in the power spectra due to true reduction in the lensing effects of $\phi$, we build simple amplitude estimates with respect to the theoretical expectations. In the absence of lensing in the data, the difference

\begin{equation}
\Delta \hat C_\ell = \debiasedChat - \hat C_\ell^{\dat}
\end{equation}
should vanish. Using the diagonal
$\sigma_\ell^2$ of the covariance matrix for the $\Delta \hat{C}_\ell$ obtained from the simulations, we form
\begin{equation}
A \equiv \frac{\sum_\ell\Delta \hat C_\ell  \Delta C_\ell^{\rm th}/\sigma_\ell^2}{\sum_\ell (\Delta C_\ell^{\rm th})^2/\sigma_\ell^2},
\label{eq:significance}
\end{equation}
using the lensed \setdeux\ simulation prediction curves (blue in Figs.~\ref{Figptt}--\ref{FigBB}, to which we add $\BnoiseOnly$) as theoretical templates $\Delta C_\ell^{\rm{th}}$, after binning. We use 40  linearly-spaced multipole bins for all spectra, except for $BB$ for which we use 20 log-linear bins. The results are shown in the top two sections of  Table~\ref{tab:table1}. The amplitudes are consistent with expectations (i.e., $A=1$) to within $2\,\sigma$. The detection significance for each case is given by the distance of $A$ from zero in units of the standard deviation $\sigma$. This can be as high as $20\,\sigma$ in $C_\ell^{TT}$. Even in $C_\ell^{BB}$, for which the lensing $B$ mode is an order of magnitude smaller than the instrumental noise (a factor of $100$ in power), there remains an approximately $3.8\,\sigma$ detection of these delensing effects using the MV estimator.
\newline
\indent
The standard deviations quoted in Table~\ref{tab:table1} are obtained within the assumption of Gaussian uncorrelated bandpowers, and represent a conservative choice of error bars. Calculating the standard deviation from the set of FFP9 simulations, the errors are systematically smaller, consistent with visible hints at a negative bin-to-bin cross-covariance observed from the delensed spectra calculated from the simulations. We use the more conservative estimates as our baseline, since given the small mismatch of the simulations to the data, we cannot guarantee that these errors bars are genuinely more accurate.
\newline
\indent
One might worry that the detection of delensing could be explained (or biased) by a bias miscalibration rather than true delensing, since the FFP9 simulation spectra do not perfectly match those of the data (see Sec.~\ref{Sims}). We can estimate the impact on the Gaussian bias of the original mismatch in power between the data and the FFP9 simulations by calculating the change in bias due to isotropic changes in power, i.e.,
\begin{equation}
\Delta B_\ell = \sum_{\ell '}\frac{\partial B_\ell}{\partial C_{\ell'}}\left (\hat{C}^{\rm dat}_{\ell'}-C^{\rm FFP9}_{\ell'} \right),
\end{equation}
where $\hat C^{\rm dat}_\ell$ are the empirical spectra of the SMICA maps and $C_\ell^{\rm FFP9}$ the average spectra of the FFP9 simulations (together with the smooth estimate of the mismatch as described in Sec.~\ref{Data}). All spectra are obtained as described in Sec.~\ref{Spectra}. The size and shape of the $\Delta B_\ell$ suggest a possible overestimation of the amplitudes $A$ by $0.1\,\sigma\: (\hat \phi^{TT})$ and $0.3\,\sigma\:(\hat \phi^{MV})$ for $C_\ell^{BB}$. In these units, the correction is comparable or smaller for the other spectra. Our detections should therefore be robust to mis-estimation of the isotropic power.
\newline
\indent 
\revision{Another possible source of error in our estimates is residual mean field in the lensing map.
The mean field is estimated by averaging lensing maps reconstructed from simulations, so an error can happen in two ways: from an insufficient number of simulations, or, more importantly, from inaccuracy of the simulations used to capture the  anisotropy of the data maps that is unrelated to lensing. In the first case, assuming perfect simulations, their finite number $N_{\rm{MC}}$ simply increases the reconstruction noise power according to
\begin{equation}
N_0 \rightarrow N_0 + \frac{C^{\phi\phi,\rm{fid}} + N_0}{N_{\rm{MC}}}.
\end{equation}
The additional noise is uncorrelated to the CMB, and causes some small degree of peak smoothing to the delensed spectra.
While it is possible to include this term in our predictions, our $N_{\rm MC} = 119$ is enough that biases on our results are at most a few percent of the error bar. The  contamination in the data bandpowers from inaccuracy of the simulations is harder to quantify, for example poor characterization of the instrument could source a slightly different mean field. The \planck\ team showed (see \citep{Planck2013CMBlensing}, Fig.~C.1) that by far the most important contribution to the mean-field power comes from masking, followed by anisotropic noise and beams (roughly two and three orders of magnitude lower, respectively). Masking is not an issue, as this is performed consistently in the data and simulations, and the remaining mean-field power is already only around $1\,\%$ or less of $N_0$ on the scales relevant for our results so likely negligible. To quantify this, we perform two simple further tests. First, since the power spectra of all the mean-field contributions are very steep functions peaking at low multipoles, we
cut multipoles of the lensing map below $\ell = 10$ (where the impact of the signal is negligible) and see no significant difference in the result. Second, we propagate a conservative $O(1)$ error in the noise mean field to our analytical predictions, using the empirical mean-field spectrum shape as template. Keeping the full range $1\leq \ell\leq 2048$, the error on the delensing amplitude is at most $0.16\,\sigma$ on our $\phi^{TT }$-delensed $\Delta C_\ell^{TT}$, and less for the other spectra.
Therefore, our detections should be robust to mean-field misestimation.
\AC{Is is worth making the simple point that if one misestimates the mean field by a fraction $\epsilon$, the additional reconstruction noise power is $\epsilon^2$ times the mean-field power?}
}
\newline
\indent
Table \ref{tab:table1} also shows ``null-hypothesis'' significances for the detection, based on errors from unlensed simulations where there is less decorrelation between the delensed and original spectra (and hence better cancellation of cosmic variance and noise). In this case the errors are smaller, giving a delensing detection significance of up to $26\,\sigma$, and $4.6\,\sigma$ in $C_\ell^{BB}$, and there is a larger improvement in the delensing detection using the MV reconstruction.
\newline
\indent
We now turn to the question of by how much did we actually delens the Planck maps. As discussed earlier, the residual lensing deflections in the maps are given by Eq.~(\ref{eq:residual}), and the parts of these residuals that arise from the reconstruction noise are, generally, not independent of the CMB. We would like to quantify the power of the residual lensing through differences between the power spectra of the delensed and original CMB. However, we cannot simply use the combination $\Delta\hat{C}_\ell -\BnoiseOnly$, as plotted in Figs.~\ref{Figptt}--\ref{FigBB}, since the debiasing subtracts \emph{all} effects of the reconstruction noise from the delensed spectra, not just the part from dependence of the reconstruction noise on the CMB fields. Instead, we add back into the delensed spectra the peak-smoothing effect of independent reconstruction noise to form $\Delta \hat C_\ell - \BnoiseOnly+\BcmbOnly$.  Here, $\BcmbOnly$ is defined in a manner analogous to $B_\ell$ and $\BnoiseOnly$ as
\begin{equation}
\BcmbOnly \equiv \left. \av{\hat C_\ell^{\rm CMB,\del} - \hat C^{\rm CMB,len}_\ell}\right|_{\text{independent}\,\hat \phi{_\clw}},
\end{equation}
where each Gaussian simulation CMB map is delensed with a lensing reconstruction obtained from a different independent Gaussian simulation (with noise and the same sky and multipole cuts as the data). \revision{This ensures, for example, that $\Delta \hat C_\ell - \BnoiseOnly+\BcmbOnly$ vanishes in the hypothetical situation that the reconstruction noise is equal in magnitude to $C^{\phi\phi}_\ell$, in which case there is no net delensing.} Note that we delens only the CMB part of the simulation to avoid contributions from delensing of the instrument noise (so the debiasing still subtracts the difference between the delensed and original noise). An alternative approach to deal with effects of delensing the noise is to consider only cross-spectra, e.g., between the two halves of the mission for Planck, for which the noise-noise terms average to zero.
However, in the case of the $TT$ and $TE$ spectra, where our delensing results are most significant, both the effects of delensing the noise and noise contributions to the biases are very small, and the benefits of using cross-spectra are minor.
We therefore chose not to pursue the cross-spectra route further here.

We use the theoretical prediction for perfect delensing, $C_\ell^{\text{fid},\rm{unl}} - C_\ell^{\text{fid},\rm{len}}$,
to fit for a relative delensing efficiency based on the full delensing simulations. In detail, we build a straightforward  $\chi^2$
\begin{equation}
\label{template}
\begin{split}
&\chi^2(\epsilon) \equiv \\
&\sum_\ell\left[ \Delta \hat C_\ell - \BnoiseOnly + \BcmbOnly - \epsilon \left (C_\ell^{\text{fid},\rm{unl}} - C_\ell^{\text{fid},\rm{len}} \right)\right]^2/\sigma^2_\ell,
\end{split}
\end{equation}
which we minimize for the efficiency $\epsilon$. The lensing efficiencies $\epsilon$ together with predictions from the FFP9 S1 simulations and reduced $\chi^2$ as goodness-of-fit statistics are presented in Table~\ref{tab:table1}. To compare with theoretical expectations we use \CAMB, to generate CMB spectra lensed with $(1- \clw_\ell)C_\ell^{\phi\phi}$ and difference these from those lensed with $C_\ell^{\phi\phi}$. We use these differences in place of $\Delta \hat C_\ell - \BnoiseOnly + \BcmbOnly$ in Eq.~\eqref{template} to compute theoretical efficiences, which are also listed in Table~\ref{tab:table1}.
Note that these efficiencies quantify the reduction in lensing power of the CMB signal, and do not account for the increase in noise power $\BnoiseOnly$ that is visible in polarization at high multipoles (see Fig.~\ref{figBiasesPred}  of Appendix~\ref{biases}). We find that the template defined in Eq.~\eqref{template}  is a consistent description of the data both in temperature and polarization, with delensing efficiencies for the acoustic features of $22\,\%$ $(\hat \phi^{TT})$ and $28\,\%$ $(\hat \phi^{\rm{MV}})$, consistent with expectations. Note that in the absence of lensing, independent reconstruction noise
would produce ``zero-point'' delensing efficiencies $\epsilon_0$, as listed in the table. They are negative since deflection noise makes the acoustic peaks smoother after delensing. We calculate $\epsilon_0$ from the same $\chi^2$ function as in Eq.~(\ref{template}), but with $\Delta \hat C_\ell - \BnoiseOnly =0$.
In all cases, the delensing efficiency is significantly larger using the MV reconstruction, but (except in the case of $BB$, where couplings are different) the significance of the amplitude measurement does not increase proportionately.
This is because with higher efficiencies the delensing effect becomes larger, and hence the lensed and delensed fields become more decorrelated, reducing the extent to which noise and cosmic variance are cancelled out when forming $\hat{C}^{\rm del}_\ell-\hat{C}^{\rm dat}_{\ell}$. We formally achieve a delensing efficiency of 7\,\% for the MV reconstruction for $C_\ell^{BB}$. For the $TT$ reconstruction the efficiency is closer to zero, meaning that the reconstruction noise has added almost exactly as much power as the delensing has removed. Nonetheless, this is consistent with a detection of delensing effects since in the absence of signal delensing the efficiency would have been negative ($\epsilon_0$).

\section{Conclusions\label{Conclusions}}
\indent
Delensing will play a prominent role in future searches for inflationary $B$-modes. We have presented the first internal delensing of CMB data, and the first demonstration of polarization delensing, using foreground-cleaned (SMICA) CMB maps from the 2015 Planck release.
\newline
\indent
The Planck lensing reconstruction is noise dominated on small scales, but has signal-to-noise close to unity around the peak of the deflection power spectrum at multipoles $\ell \approx 60$. The dominant lensing smoothing effect on the temperature and $E$-mode polarization power spectra comes from lensing modes that are larger than the acoustic scale, peaking at $\ell \sim 150$ but with a significant dependence down to larger scales (as shown in Fig.~\ref{Figeps}). We therefore achieve a good delensing efficiency, close to 30\,\%, both in temperature and $E$-mode polarization, with detection significances of at least $18\,\sigma$. This compares well with the temperature delensing efficiencies achieved using CIB delensing by Ref.~\cite{Larsen:2016wpa}; the CIB is a much higher signal-to-noise tracer of the lensing potential on small scales, but is hard to separate from Galactic dust emission on large-scales that contribute significantly to the smoothing effect (and where the sensitivities of lensing reconstructions with Planck are highest).

We also achieve an approximately 7\,\% delensing efficiency in the $BB$ power spectrum (where 10\% was expected), with the delensing effects detected at $4.5\,\sigma$. For the $B$-mode polarization, the lensing signal is all produced by lensing of $E$ modes, and couples with a much broader kernel of lensing modes out to smaller scales; the delensing efficiency is therefore correspondingly lower than for the smoothing effect on $TT$, $TE$ and $EE$. CIB-delensing has the potential to perform much better on this data set. It is worth noting that we detect this reduction in lensing $B$ modes with high significance because we are targeting differences in the spectra where instrumental noise cancels out. At Planck noise levels, \emph{direct} detection of the power in lensing $B$ modes themselves is not possible at high significance. From the point of view of detecting primordial $B$ modes, the relevant spectrum to consider is $C^{BB,\rm{del}}_\ell + N_\ell^{BB}$, where $N_\ell^{BB}$ is the angular power spectrum of the instrumental noise. For Planck, $N_\ell^{BB}$ is larger than the lens-induced $B$-mode power, $C^{BB,\rm{len}}_\ell$, by at least a factor of 100, so reducing the lensing power by 7\,\% gives almost no improvement in constraints on primordial $B$-mode polarization.
Neither do we want to claim that our method is well suited for lensing $B$-mode detection: previous work using the same Planck data cross-correlated a $B$-mode template built from $\hat \phi $ and the $E$ polarization and found detections at rather higher significances of $10\,\sigma$ (\PlanckLens) and $12\,\sigma$ \cite{Ade:2015nch}.
\newline
\indent
We have shown that our results are robust to a series of possible systematic effects in the Planck data. They are consistent with two independent sets of predictions: one based on a simple theoretical calculation using \CAMB, and  one based on simulations that captures some of the real-life complications such as sky masking, anisotropic noise and beams, etc. The main difficulty lies in the biases that enter the delensed spectra, which at Planck sensitivity must be accurately modelled since they are of comparable amplitude to the signal. These biases arise from non-independence of the lensing potential reconstruction noise and the CMB maps. Reference~\cite{Green:2016cjr} recently developed non-perturbative models for the delensed temperature and polarization spectra, but only accounted for independent lensing noise and so did not identify these additional biases.
We chose to subtract the (dependent) biases using Gaussian simulations, but showed that they can also be well understood using a perturbative analysis. While the importance of these biases should decrease for the upcoming CMB experiments, which will produce lens reconstructions with much higher signal-to-noise, it is likely to remain important to model them.

More generally, while our current methodology is sufficient for a strong detection of delensing, not all aspects are optimized and there is likely room for further improvements. For example, a methodology with less dependence on the instrumental noise, such as, e.g., cross-spectra, or optimized multipole cuts, would be desirable, lessening the need for Monte-Carlo simulations and reducing the requirements on their accuracy. Furthermore, remapping filtered versions of the CMB maps may improve optimality, as discussed in Ref.~\cite{Green:2016cjr}. Also, a full analysis of the covariance of the delensed spectra is left for future work. This is likely non-trivial given the dependence of the reconstruction noise on the CMB fields, but is important to meet the goal of correctly extracting all information from the delensed power spectra.

\subsection*{Acknowledgements}
We thank Duncan Hanson for making his lensing simulation and reconstruction libraries available, upon which part of this work is based. We also thank Patricia Larsen for helpful discussions at an early stage of this work. \JCorcid\ and \ALorcid\ acknowledge support from the European Research Council under
the European Union's Seventh Framework Programme (FP/2007-2013) / ERC Grant Agreement No. [616170],
and \ALorcid\ and AC from the Science and Technology Facilities Council {[}grant numbers ST/L000652/1 and ST/N000927/1, respectively{]}. This research used resources of the National Energy Research Scientific Computing Center, a DOE Office of Science User Facility supported by the Office of Science of the U.S. Department of Energy under Contract No. DE-AC02-05CH11231.
\appendix

\newcommand{\beq}{\begin{equation}}
\newcommand{\enq}{\end{equation}}
\newcommand{\vecr}{\vr}

\section{Delensing biases}
\newcommand{\uncorrb}{B^{(\textrm{indep.})}_\ell}
\newcommand{\corrb}{B^{(\textrm{dep.})}_\ell}
\newcommand{\uncorrbr}{B^{(\textrm{indep.})}}
\newcommand{\corrbr}{B^{(\textrm{dep.})}}
\label{biases}

In this appendix we give an analytical derivation of the main contributions to the bias in the delensed power spectra for the case of an isotropic survey (i.e., full sky with isotropic noise). We adopt a real-space approach, with starting point the $T$, $Q$ and $U$ maps. This allows a streamlined derivation and efficient numerical implementation,
both for the temperature-based or the minimum-variance estimator that we use in this work. On the other hand, the drawback is that we do not offer a formula for the biases for arbitrary pairs of estimators such as $TE$, $EE$, $EB$ and $TB$.
We calculate the leading contributions to the bias from the lensing reconstruction noise, including the effect of the statistical dependence between this noise and the CMB maps.
\newline
\indent We proceed as follows. We first obtain the real-space two-point functions of the delensed $T$, $Q$ and $U$ fields. For the terms we are interested in, they can all be written in simple forms involving at most products and convolutions of real-space homogeneous two-point functions that can be computed efficiently with simple harmonic transforms. After these are combined the inverse harmonic transform gives the delensed $T$, $Q$ and $U$ (anisotropic) spectra and cross-spectra. Finally, these are rotated to obtain the delensed $T$, $E$ and $B$ spectra. We work in the flat sky approximation throughout, with

\begin{equation}\label{QU2EB}
\begin{split}
E(\vell) &= Q(\vell) \cos(2\psi_{\vell}) + U(\vell) \sin(2\psi_{\vell}) \\
B(\vell) &= -Q(\vell) \sin(2\psi_{\vell}) + U(\vell) \cos(2\psi_{\vell}) .
\end{split}
\end{equation}
Here, the wavevector $\vell$ has Cartesian components $l(\cos\psi_\vell,\sin\psi_\vell)$.
In what follows, early Latin indices (e.g., $a$, $b$) run over the two axes of the flat sky, $i$, $j$ etc. run over the full set of maps $T$, $Q$ and $U$, and Greek letters (e.g., $\alpha$, $\beta$) run over the subset of these maps that is used for the lensing quadratic estimator (i.e., either $T$ alone or $T$, $Q$ and $U$ for the MV estimator).
\newline
\indent
Let $X^i(\vx)$ be one of the data maps $T$, $Q$ or $U$, inclusive of noise. We write the two-point function (statistically homogeneous, but anisotropic) as
\begin{equation}
C^{ij}(\vecr) \equiv \av{X^i(\vx) X^j(\vy)} ,
\end{equation}
where $\vr \equiv \vx - \vy$. These two-point functions are easily calculated from the $T$, $E$ and $B$ spectra and cross-spectra using the relations in Eq.~\eqref{QU2EB}.
We shall also need correlators of derivatives, e.g.,
\begin{align}
\av{ \nabla_a X^i(\vx) \nabla_b X^j(\vy)} &= -\av{ \nabla_a \nabla_b X^i(\vx) X^j(\vy)} \nonumber \\
&= -C^{ij}_{,ab}(\vecr) ,
\label{eq:corrderiv}
\end{align}
which are also obtained easily in Fourier space.
In real space, our definition of the bias $B_\ell$ [see Eq.~\eqref{Bdef}]
becomes
\beq
B^{ij}(\vr) =  \left. \av{X^{i,\rm del}(\vx) X^{j,\rm del}(\vy)-X^i(\vx) X^{j}(\vy)} \right|_{\text{Gaussian CMB}},
\enq
which is evaluated for Gaussian CMB fields with the correct lensed spectra. Equivalently, this bias can be thought of as the expected difference between the delensed spectra and the original spectra when only (disconnected) Gaussian contractions of the CMB fields are included in the calculation of the expectation values. The true expected differences are the sum of the bias, plus the connected contribution to $\av{X^{i,\rm del}(\vx) X^{j,\rm del}(\vy)}$ from the 4-point (and higher) connected moments of the lensed CMB fields. It is the latter that describes the reduction of real lensing effects in the spectra.

In this appendix, we calculate the bias and the connected contribution perturbatively
by expanding to second order in the applied displacement $\nabla \hat \phi$, so we can write at each point
\begin{equation}
\label{Expand}
X^{i,\rm del} = X^i - \nabla_a \hat \phi \nabla^a X^i +\frac 12  \nabla_a \hat \phi  \nabla_b \hat \phi\nabla^a\nabla^b X^i.
\end{equation}
It follows that
\begin{equation}
\begin{split}
&\av{X^{i,\rm del}(\vx) X^{j,\rm del}(\vy)-X^i(\vx) X^{j}(\vy)} = \\
&- \av{X^i(\vx)(\nabla_a \hat{\phi} \nabla^a X^j)(\vy)} \\
&+ \frac{1}{2} \av{(\nabla_a \hat{\phi} \nabla^a X^i)(\vx)
(\nabla_b\hat{\phi} \nabla^b X^j)(\vy)} \\
&+ \frac{1}{2} \av{X^i(\vx) (\nabla_a \hat{\phi}\nabla_b\hat{\phi} \nabla^a \nabla^b X^j)(\vy)}  \\
&+  (i,\vx \leftrightarrow j,\vy) .
\label{eq:corrdiff}
\end{split}
\end{equation}
We first discuss the connected contribution to the right-hand side, putting the discussion of Sec. \ref{Spectra} of our expectations on firm grounds, before considering the bias $B$. For the latter, we distinguish two contributions. The dominant one originates from the statistical dependence between the reconstruction noise and the CMB fields, and involves Gaussian contractions between the CMB fields in $\hat{\phi}$ and $X^i$ and $X^j$. We then discuss the sub-dominant contribution to the bias, which is the additional lensing-like power due to the reconstruction noise. Mathematically, this contribution arises from terms where the CMB fields in the reconstruction are correlated across pairs of $\hat{\phi}$.

\subsection{Connected contribution}
The terms on the right hand side of Eq.~\eqref{eq:corrdiff} involve the 4- and 6-point functions of the lensed
CMB.
In the first term, the largest connected contribution will come from the primary coupling of
the trispectrum (see~\cite{Hu:2001fa,Hanson:2010rp}), which here, at leading order, is equivalent to taking the expectations of $\nabla_a \hat{\phi}(\vy)$ and $X^i(\vx)\nabla^a X^j(\vy)$
over the \emph{unlensed} CMB at fixed $\phi$ before averaging their contraction over realizations of $\phi$, i.e.,
\begin{multline}
\av{X^i(\vx)(\nabla_a \hat{\phi} \nabla^a X^j)(\vy)}_c^{\text{primary}}  = \\
\av{\langle \nabla_a \hat{\phi}(\vy) \rangle_{\text{CMB}} \av{X^i(\vx)\nabla^a X^j(\vy)}_{\text{CMB}}}_\phi .
\end{multline}
For our Wiener-filtered potential estimate, $\langle \hat{\phi} \rangle_{\text{CMB}} = \mathcal{W} \star \phi$. We also define the correlator of the gradient of the Wiener-filtered potential with the gradient of the potential,
\begin{align}
C^{ab}_{[\mathcal{W}\phi]\phi}(\vr) &= \av{\nabla^a [\mathcal{W}\phi](\vx) \nabla^b \phi(\vy)} \nonumber \\
&= \int \frac{\ud^2 \vell}{(2\pi)^2}\,  (i\ell_a) (-i\ell_b)\clw_\ell C_\ell^{\phi\phi} e^{i \boldsymbol{\ell}\cdot \vr} ,
\end{align}
and $C^{ij,\text{unl}}_{,ab}(\vr)$, the equivalent of Eq.~\eqref{eq:corrderiv} for noise-free unlensed fields (and ignoring beam effects). Substituting the leading term in the series expansion for the lensed fields $X^i \approx X^{i,\rm{unl}} + \nabla_c \phi \nabla^c X^{i,\rm{unl}}$ we then find
\begin{multline}
\av{X^i(\vx)(\nabla_a \hat{\phi} \nabla^a X^j)(\vy)}_c^{\text{primary}} \approx
-C^{ij,\text{unl}}_{,ab}(\vr)\\
\times \left[C^{ab}_{[\mathcal{W}\phi]\phi}(\vr) -C^{ab}_{[\mathcal{W}\phi]\phi}(\boldsymbol{0}) \right] \, .
\end{multline}
The additional contribution from $(i,\vx \leftrightarrow j,\vy)$ is the same.
In harmonic space, this is exactly the usual expression for the change in the power spectrum due to lensing at lowest perturbative order~\cite{Hu:2000ee}, but with the $C_\ell^{\phi\phi} \rightarrow \mathcal{W}_\ell C_\ell^{\phi\phi}$.

The remaining expectation values on the right of
Eq.~(\ref{eq:corrdiff}) involve the 6-point
function of the lensed CMB fields. The 6-point function can be written as a sum of the
connected 6-point function, products of the connected 4-point
functions and the 2-point functions, and the fully-disconnected part
involving products of three 2-point functions (which contributes only to the bias).
Working to
$O(C_\ell^{\phi\phi})$, we can ignore the connected 6-point function~\cite{Kesden:2002jw}.
For those parts involving the connected 4-point functions, we expect that
the dominant contributions will come from the most tightly-coupled
terms (i.e., those that factor most under the reconstruction weights
in $\hat{\phi}$~\cite{Hanson:2010rp}), which will therefore involve $\langle \hat{\phi}
\hat{\phi} \rangle_c$. Keeping only the primary couplings, we have, for
example,
\begin{multline}
\av{(\nabla_a \hat{\phi} \nabla^a X^i)(\vx)
(\nabla_b\hat{\phi} \nabla^b X^j)(\vy)}_c \approx \\
\av{\nabla_a [\mathcal{W}\phi](\vx) \nabla_b [\mathcal{W}\phi](\vy)}
\av{\nabla^a X^i(\vx) \nabla^b X^j(\vy)} ,
\end{multline}
which we can write as $- C^{ij}_{,ab}(\vr) C^{ab}_{[\mathcal{W}\phi][\mathcal{W}\phi]}(\vr) $, where $C^{ab}_{[\mathcal{W}\phi][\mathcal{W}\phi]}(\vr)$ is the two-point function of the gradient of the Wiener-filtered potential. A similar calculation for the final term in Eq.~\eqref{eq:corrdiff} gives its dominant contribution as $C^{ij}_{,ab}(\vr) C^{ab}_{[\mathcal{W}\phi][\mathcal{W}\phi]}(\boldsymbol{0})$.

Combining the above results, we find the dominant connected contribution
\begin{multline}
\av{X^{i,\rm del}(\vx) X^{j,\rm del}(\vy)-X^i(\vx) X^{j}(\vy)} \supset \\
2 C^{ij,\text{unl}}_{,ab}(\vr) \left[C^{ab}_{[\mathcal{W}\phi]\phi}(\vr) -C^{ab}_{[\mathcal{W}\phi]\phi}(\boldsymbol{0}) \right] \\
-  C^{ij}_{,ab}(\vr)\left[C^{ab}_{[\mathcal{W}\phi][\mathcal{W}\phi]}(\vr)
- C^{ab}_{[\mathcal{W}\phi][\mathcal{W}\phi]}(\boldsymbol{0}) \right] .
\end{multline}
At leading order in $C_\ell^{\phi\phi}$, we can replace $C^{ij}_{,ab}(\vr)$ in the final term with the sum of the unlensed 2-point function $C^{ij,\text{unl}}_{,ab}(\vr)$ and the noise contribution to $C^{ij}_{,ab}(\vr)$. Furthermore, we can write
\begin{multline}
\av{X^i(\vx) X^j(\vy)} \approx C^{ij,\text{unl}}(\vr) \\
- C^{ij,\text{unl}}_{,ab}(\vr) \left[C^{ab}_{\phi\phi}(\vr) -C^{ab}_{\phi\phi}(\boldsymbol{0}) \right] + \text{noise} ,
\end{multline}
where $C^{ij,\text{unl}}(\vr)$ is the 2-point function of unlensed, noise-free fields and the final term is the 2-point function of the instrument noise. It follows that
\begin{multline}
\av{X^{i,\rm del}(\vx) X^{j,\rm del}(\vy)} = C^{ij,\text{unl}}(\vr) \\
- C^{ij,\text{unl}}_{,ab}(\vr) \left[C^{ab}_{[(1-\mathcal{W})\phi][(1-\mathcal{W})\phi]}(\vr) -C^{ab}_{[(1-\mathcal{W})\phi][(1-\mathcal{W})\phi]}(\boldsymbol{0}) \right] \\+ \text{noise} .
\end{multline}
Here, the noise contribution is the 2-point function of the noise lensed by $\mathcal{W}\star \phi$, as if the noise were displaced by $-\nabla (\mathcal{W}\star \phi)$. For the signal contribution, we see that the delensed 2-point function is as if the unlensed CMB were displaced by $\nabla(\phi - \mathcal{W}\star \phi)$. These results are consistent with the intuitive discussion in Sec.~\ref{Spectra}.

\subsection{Bias due to dependency of the reconstruction noise on the CMB}
\label{corrb}
The main contribution to the bias comes from the Gaussian (disconnected) contribution to the first term on the right of Eq.~\eqref{eq:corrdiff}:
\begin{equation} \label{corrbb}
\corrbr_{ij}(\vecr) =\av{-X^i (\vx) \nabla_a \hat \phi(\vy) \nabla^a X^j(\vy)}_G + (i,\vx \leftrightarrow j,\vy) ,
\end{equation}
where the subscript $G$ denotes that only Gaussian contractions of the CMB fields are included. The only such contractions are between the CMB fields in $\hat{\phi}$ and the $X^i$ and $X^j$ (provided that the mean-field has been accurately subtracted from $\hat{\phi}$), and so this bias would vanish if the lensing reconstruction were independent of the CMB. For Planck, this term is the dominant source of bias for all spectra in the case of $\hat \phi^{\rm{MV}}$-delensing, and for the $TT$ spectrum for $\hat \phi^{TT}$-delensing.

We proceed as follows. A filtered quadratic estimator such as $TT$ or MV has separable weights and can always be written as
\beq
\label{qest}
\nabla_a \hat \phi(\vx)  = \int \frac{\ud^2 \vz}{2\pi} \: F^{ab}(\vx-\vz)[V^{\alpha \beta} X^{\alpha}](\vz)[W_b^{\gamma \beta} X^{\gamma}](\vz) ,
\enq
with implicit summation over repeated indices. The operation $V^{\alpha\beta}X^\alpha$ performs the inverse filtering of the CMB maps, while $W_b^{\gamma\beta}X^\gamma$ returns the gradient of the Wiener-filtered fields.
For example, in the case of the $TT$ estimator, $V$ and $W_b$ are given in harmonic space by
\begin{align}
V^{TT}(\vell) &= \frac{b_\ell}{b_\ell^2C_\ell^{TT} + N_\ell^{TT}}, \\
W_{b}^{TT}(\vell) & =  i\ell_b  \frac{b_\ell C_\ell^{TT} }{b_\ell^2C_\ell^{TT} + N_\ell^{TT}}.
\end{align}
Here, $b_\ell$ is a fiducial estimate of the transfer function (e.g., due to the beam smoothing) of the temperature map and $N_\ell^{TT}$ a similar fiducial estimate of the noise spectrum.
The filtering $F$ has three types of contributions: (1) a normalisation $A_\ell$;\footnote{
For optimal inverse-covariance filtering of the CMB fields, the normalisation is simply the Gaussian reconstruction noise power $N_{\ell,0}$. However, for practical reasons, in this work we ignore the $TE$ correlation when filtering, following~\PlanckLens.  In this case, the normalisation is no longer exactly equal to $N_{\ell,0}$.}
(2) the projection of the Cartesian components of the estimates of the deflection field 
onto gradient or curl components; and (3), any a posteriori filtering, such as the Wiener filtering we adopt in this work. In harmonic space,
\beq
F^{ab}(\vell) =(i \ell_a) \clw_\ell  (-i\ell_b) A_\ell.
\enq
\newline
\indent
Forming the Gaussian contractions in Eq.~\eqref{corrbb}, using the reconstruction~\eqref{qest} results in
\beq
\begin{split}
&\corrbr_{ij}(\vecr) = \\
&\int \frac{\ud^2 \vz}{2\pi}\:  F^{ab}(\vz) [V^{\alpha \beta}C^{\alpha i}_{,a}](-\vz) [W_b^{\gamma\beta}C^{\gamma j}](\vecr - \vz) \\
+&\int \frac{\ud^2 \vz}{2\pi} \: F^{ab}(\vz) [V^{\alpha \beta}C^{\alpha j}](\vr-\vz) [W_b^{\gamma \beta}C^{\gamma i}_{,a}](- \vz) \\
+&(i,\vecr \leftrightarrow j,-\vecr)
\end{split}
\label{eq:corrbias}
\enq
All the terms can be evaluated with 2D FFT techniques.
\newline
\indent
To understand the form of the bias, consider the simple case of the temperature reconstruction delensing the $TT$ spectrum.
Neglecting instrument noise and beams, the correlated bias term given in Eq.~\eqref{eq:corrbias} can be written in harmonic space using the flat-sky approximation as
\begin{multline}
B_\ell^{\rm (dep.)} = 4\int \frac{{\rm d}^2\vell'}{(2\pi)^2} \vell'\cdot(\vell-\vell') \\\quad\times N^{TT}_{\ell',0} \clw_{\ell'} g^{TT}(\vell, \vell')  C^{TT}_\ell C^{TT}_{|\vell-\vell'|},
\label{eq:corrbiasharmonic}
\end{multline}
where the temperature quadratic estimator weight function
$g^{TT}$ is given by
\begin{equation}
g^{TT}(\vell,\vell') \equiv \frac{ \vell'\cdot \vell C^{TT}_{\ell} +
 \vell'\cdot (\vell'-\vell)  C^{TT}_{|\vell'-\vell|}}{2  C^{TT}_{\ell} C^{TT}_{|\vell'-\vell|}} .
\end{equation}
To see the qualitative effect, we series expand the integrand in Eq.~\eqref{eq:corrbiasharmonic} for $\ell'\ll \ell$, i.e., modelling only large-scale lensing reconstruction modes. Assuming the fiducial model, doing the angular integral and keeping the leading term, we have
\begin{multline}
\frac{\ell^2B_\ell^{\rm (dep.)}}{2\pi}\sim - \left( \frac{3}{2}\frac{\ud^2 D_\ell}{\ud\ln\ell^2} + \frac{\ud D_\ell}{\ud \ln\ell}\right)  \\
\times\frac{1}{4}\int_{\ell'\ll\ell} \frac{\ell' \ud\ell'}{2\pi} \ell'{}^4N^{TT}_{\ell',0} \clw_{\ell'},
\label{eq:squeezedbias}
\end{multline}
where $D_\ell = \ell(\ell+1)C^{TT}_\ell/2\pi$. The first term describes the $\ell$-dependence and the integral determines the amplitude.
This can be compared to the effect of lensing on the CMB power spectrum in the limit of large lenses\footnote{
The leading eigenvector that determines the lensing effect on the $TT$ power shown in Fig.~\ref{Figeps} is reasonably well approximated by $\ell v_\ell \propto \ell^6 C_\ell^{\phi\phi} e^{-(\ell/225)^2}$, corresponding to the
contribution per $\log \ell$ to $\la \kappa_{\ell'\ll\ell}^2\ra \approx \la \kappa_s^2\ra$, where $\kappa_s$ is the convergence smoothed on the characteristic scale of the CMB acoustic peaks.
}~\cite{Bonvin:2015uha,Lewis:2016tuj}
\begin{equation}
D_\ell^{\rm unl}-D_\ell^{\rm len} \approx
-\left( \frac{3}{2}\frac{\ud^2 D_\ell}{\ud\ln \ell^2} + \frac{\ud D_\ell}{\ud \ln \ell}\right)\frac{\la \kappa_{\ell' \ll \ell}^2\ra}{2},
\label{eq:squeezedCL}
\end{equation}
where the second derivatives give the characteristic lensing smoothing effect.
Here, $\langle \kappa^2_{\ell'\ll\ell}\rangle$ is the mean-squared convergence of large-scale lensing modes.
The $\ell$-dependence of Eq.~\eqref{eq:squeezedbias} has the same form, also giving a smoothing effect, so the correlated bias term has approximately the same sign and form as the lensing signal itself (as argued qualitatively in the introduction, and can be seen numerically in Fig.~\ref{Figptt}).
Note that $\frac{1}{4}\ell^4N^{TT}_{\ell,0}$ is just the reconstructed convergence noise spectrum, so that the amplitude of the bias (set by the integral in Eq.~\ref{eq:squeezedbias}) is given by the variance of the filtered convergence reconstruction noise, in the same way that the convergence variance determines the amplitude of the lensing effect in Eq.~\eqref{eq:squeezedCL}.


The size of the correlated bias term depends on the form of the Wiener filter at high $\ell$, since the integral in Eq.~\eqref{eq:squeezedbias} is not rapidly converging. For example, the size of the bias can be made a factor of two smaller by setting the filter to zero at $\ell > 150$, with only a small decrease in delensing efficiency for $T$ and $E$.

\begin{figure*}[h!tp]
\includegraphics[width = 0.8\textwidth, trim=0cm 5cm 0cm 5cm, clip]{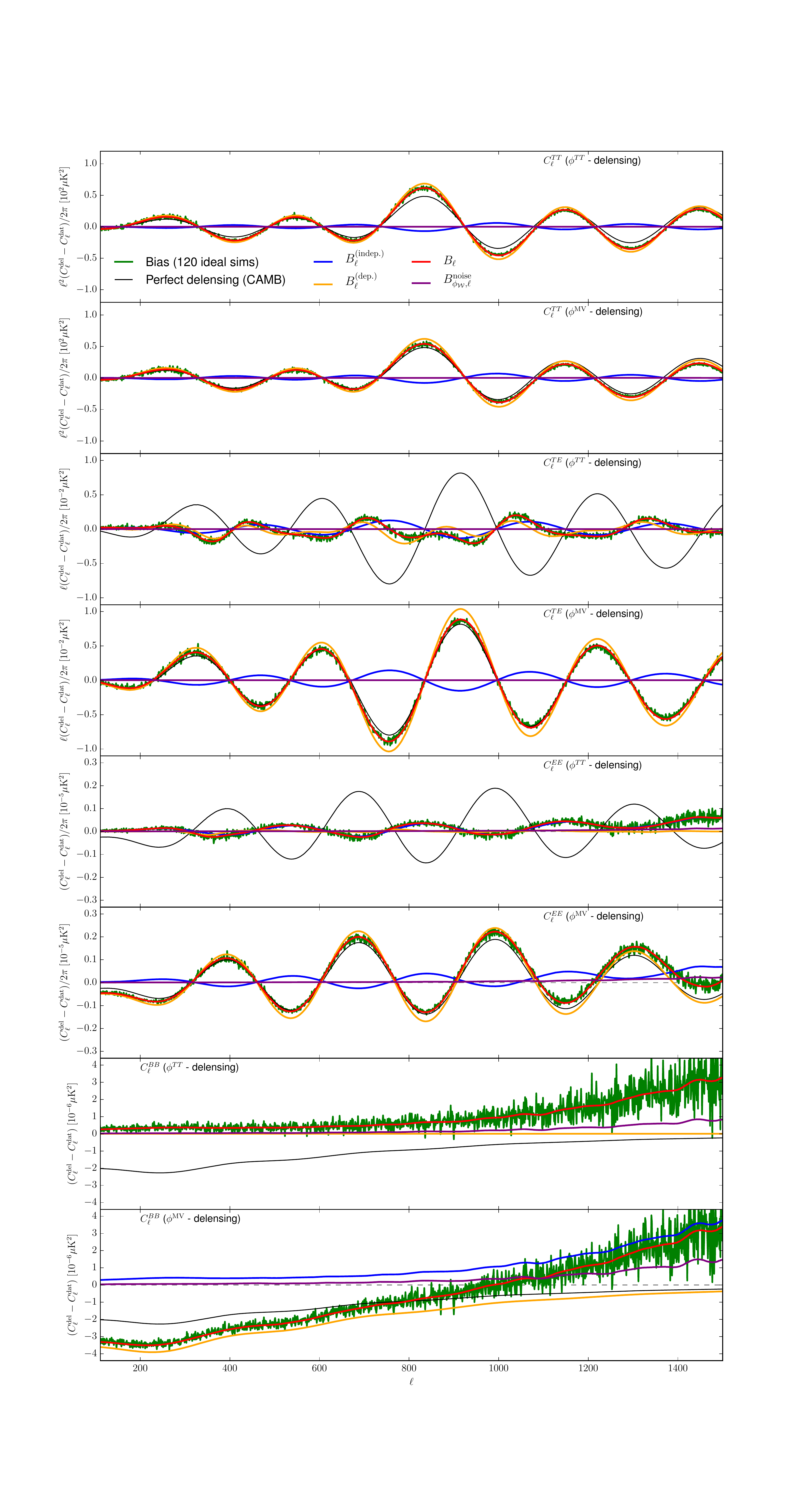}
\caption{\label{figBiasesPred}  Perturbative prediction for the two bias terms $\corrb$ (orange) and $\uncorrb$ (blue), together with their sum $B_\ell$ (in red). Also shown in green is the non-perturbative bias measured on 120 idealized Planck-like simulations. Band-powers (not shown), calculated in the same way as in the main text, are in perfect agreement given the Monte-Carlo noise in the simulation measurements.
Also shown is the noise delensing bias $\BnoiseOnly$, defined in Eq.~\eqref{Bnoise}, in purple.
}
\end{figure*}

\subsection{Bias from lensing with quadratic estimator noise}
 \label{uncorrb}
We now consider the bias, $\uncorrbr$, from the additional lensing-like power of the reconstruction noise. This arises from Gaussian contractions between the CMB fields across pairs of $\hat{\phi}$, so only the last two terms in Eq.~\eqref{eq:corrdiff} contribute.
Let $N_{ab}(\vecr) = \av{\nabla_a \hat \phi(\vx) \nabla_b \hat\phi(\vy) }_G$ be the two-point function of the displacement estimate for Gaussian CMB fields.
In our case, it contains the Wiener-filtered reconstruction noise spectrum, i.e.,
\beq
N_{ab}(\vecr)  =
\int \frac{\ud^2 \vell}{(2\pi)^2}\,  (i\ell_a) (-i\ell_b)N_{\ell,0} \clw_\ell^2 e^{i \boldsymbol{\ell}\cdot \vr}.
\label{eq:twopointrecon}
\enq
We can write $\uncorrbr$ as
\begin{equation}
\begin{split}
\uncorrbr_{ij}(\vecr) = -C^{ij}_{,ab}(\vecr) \left [N_{ab}(\vecr)  -N_{ab}(\boldsymbol{0})  \right].
\label{eq:uncorrbias}
\end{split}
\end{equation}
This bias is simply the convolution in Fourier space of the displacement spectrum with the second derivative of the two-point function of the CMB fields.
In harmonic space, Eq.~\eqref{eq:uncorrbias} is of the form of the usual expression for the lensed power spectrum to lowest perturbative order~\cite{Hu:2000ee}, where here the (de)lensing is operating on the observed field, and the deflections are the filtered lensing reconstruction noise.
The bias $\uncorrbr$ therefore corresponds to an additional lensing-noise smoothing of the delensed field, decreasing the difference the between the delensed field and the observed lensed field around the acoustic peaks.
The noise delensing correction $\BnoiseOnly$, defined by Eq.~\eqref{Bnoise} in the main text, is given by a similar expression to $\uncorrbr$ but with $C^{ij}$ replaced by the 2-point function of the instrument noise and $N_{\ell,0}$ replaced by $C^{\phi\phi}_\ell$ in Eq.~\eqref{eq:twopointrecon}.

\subsection{Comparison with simulations}
\indent Figure~\ref{figBiasesPred} shows the two bias terms for the full set of spectra $C_\ell^{TT}$, $C_\ell^{TE}$, $C_\ell^{EE}$ and $C_\ell^{BB}$, both for $\hat\phi^{TT}$ and $\hat\phi^{\rm{MV}}$-delensing, computed as described in this appendix. Their sum is compared to the empirical measurement of the total, non-perturbative bias from 120 idealized simulations from the set \settrois. Also shown is the noise delensing contribution $\BnoiseOnly$. The agreement with the simulations is everywhere very good, and is, in fact, limited by the Monte-Carlo noise from the finite number of simulations. Note that the MV estimator has lower reconstruction noise than the $TT$ estimator, but the \emph{filtered} noise variance actually increases as the noise goes down, so the independent bias is slightly larger in the case of MV.
We note that the shape of the biases shown in Fig.~\ref{Figp} are qualitatively different for the MV case to those in Fig.~\ref{figBiasesPred}. These differences are due to the
independent filtering of Planck temperature and polarization (i.e., setting $C_\ell^{TE}$ to zero in the filtering) that we adopt in the main text, while the results in Fig.~\ref{figBiasesPred} are for the genuine minimum variance estimator without this assumption.

\section{Impact of optimal lens reconstruction on delensing efficiency \label{Iterative}}
\newcommand{\Cov}{\rm{Cov}}

\begin{figure}[htp]
\includegraphics[width = 0.49\textwidth]{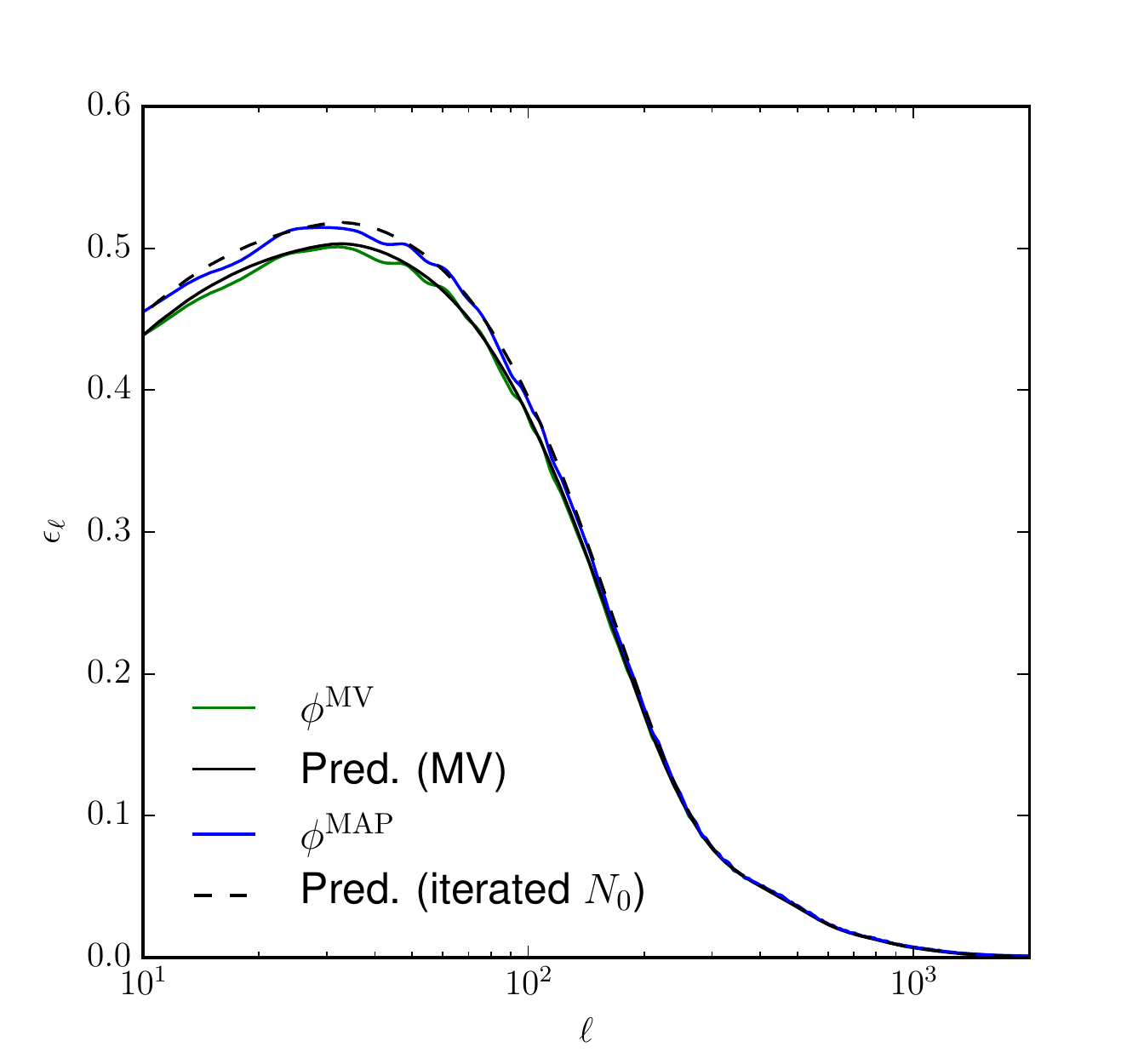}
 \caption{ \label{fig:epsit}
Delensing efficiencies for the maximum a posteriori (MAP) potential reconstruction $\hat \phi^{\rm{MAP}}$ (blue) [See Eq.~\eqref{effMAP}], in comparison to the minimum variance (MV)  quadratic estimator (green). The blue curve was obtained averaging over 64 MAP reconstructions of flat-sky Planck-like simulations as described in the main text. We use the Wiener-filtered MV estimator as a starting point for the iterative procedure leading to the MAP solution.
Also shown is the heuristic analytic prediction obtained by calculating reconstruction noise levels and delensed spectra iteratively (black, dashed), and the prediction for the MV estimator, the Wiener filter $\mathcal W_\ell$ (black, solid) [Eq.~\eqref{WF2}].}
\end{figure}

In this appendix we consider the potential improvements from delensing with a more optimal lensing reconstruction than the simple quadratic estimators used in the main text. We can estimate the expected improvements heuristically with an iterative scheme as follows: first, we calculate an improved reconstruction noise using theory spectra delensed using Eq.~\eqref{resLP}; second, we calculate new delensed spectra for this improved reconstruction noise; then iterating this procedure until convergence~\cite{Smith:2010gu}. We obtain improvements in $N_{\ell,0}$ of 7.5\,\% ($TT$) and 8.5\,\% (MV) around the peak of the lensing spectrum, decaying sharply thereafter. Since $N_{\ell,0}$ has similar magnitude to $C_\ell^{\rm{fid},\phi\phi}$ we might expect the delensing efficiencies to increase by around 2\,\%.
\newline
\indent
To test whether such iterative delensing would significantly change our results, we use an optimal lensing potential estimator on the flat-sky simulations \settrois. Details of the implementation will be presented elsewhere \cite{Carron:2017mqf}; here, we just give a brief summary description.
\newline
\indent
We model the signal $X = T,Q,U$ observed in pixel $\vx_i$ as the convolution of the deflected CMB with an effective beam function $B$, to which we add independent noise
\begin{equation}
X^{\rm dat}_i = \int \ud^2 \vy \: B(\vx_i - \vy) X^{\mathrm{unl}}(\vy + \deflect(\vy)) + n_i.
\end{equation}
Unlensed CMB fields, and the noise in each pixel, obey Gaussian statistics. The likelihood of the data for a given fixed deflection field is therefore also Gaussian. The pixel-pixel covariance can be written in compact notation using a series of linear operators as follows
\begin{equation}
\label{anisoCov}
\left [ \mathrm{Cov}_{\deflect}\right]_{ij} \equiv \av{X^{\rm dat}_iX^{\rm dat}_j} = \left[ B \:  C_{\deflect} \: B^t \right]_{ij} + N_{ij},
\end{equation}
where $N$ is the noise covariance.
The signal covariance matrix $C_{\deflect}$ is given in position space by
\begin{equation}
\begin{split}
C_{\deflect}(\vx,\vy) &\equiv \av{X^{\mathrm{unl}}(\vx +\deflect(\vx) )X^{\mathrm{unl}}(\vy +\deflect(\vy) )} \\ &= C^{\rm{unl}}(\vx +\deflect(\vx)-\vy -\deflect(\vy)).
\end{split}
\end{equation}
Under this assumption of Gaussian unlensed CMB and noise, and further assuming a pure gradient deflection field $\deflect = \nabla \phi$, the log-posterior probability density for the lensing potential can be written as (ignoring $\phi$-$T$ and $\phi$-$E$ cross-correlations)
\begin{equation}
\begin{split}
&\ln p(\phi | X^{\rm dat}) = \\
&-\frac{1}{2} \left(X^{\rm dat}\right)^t\mathrm{Cov}_\phi^{-1} X^{\rm dat} - \frac{1}{2} \ln \det \mathrm{Cov}_\phi - \frac{1}{2}\sum_{\vell} \frac{|\phi_\vell |^{2}}{C^{\rm{fid},\phi\phi}_\ell}.
\end{split}
\end{equation}
Here, $\phi_\vell$ is the discrete Fourier transform of the pixelized field $\phi$.
We implemented a quasi-Newton minimizer of this posterior PDF, sharing some similarities to the method of Refs.~\cite{HirataSeljak2003a,HirataSeljak2003b} (but involving no approximations) on the flat sky. This uses the full set of temperature and polarization maps.  The Wiener-filtered, MV quadratic estimate of the potential map serves as a starting point, and gradient and curvature information is then used to find the maximum a posteriori (MAP) point, defined through
\begin{equation}
\label{effMAP}
\hat\phi^{\rm{MAP}} =\underset{\phi}{\rm{arg max}}\ln p(\phi | X^{\rm dat}).
\end{equation}
We obtain the MAP solution $\hat \phi^{\rm{MAP}}$ on 64 flat-sky simulations from \settrois. From these solutions we calculate a delensing efficiency from the squared cross-coefficient to the input potential $\phi^{\rm{in}}$ [See Eq.~\eqref{CorrEps}],
\begin{equation}\label{effMAP}
\epsilon_\ell^{\rm{MAP}} = \frac{\left (C^{\phi^{\rm{MAP}} \phi^{\rm{in}}}_\ell \right)^2}{C_\ell^{\phi^{\rm{MAP}}\phi^{\rm{MAP}}} C_\ell^{\phi^{\rm{in}}\phi^{\rm{in}}}}.
\end{equation}
Figure~\ref{fig:epsit} shows $\epsilon_\ell^{\rm MAP}$ from the 64 simulations as the blue curve, together with the quadratic MV estimator efficiency (green). We find that the
MAP estimator's efficiency is fairly accurately predicted by the heuristic procedure described above (dashed). The improvement of the quadratic estimator is small, and is nowhere better than a few percent.
We see no qualitative change in the (biased) delensed spectra on the simulations, and only relatively small quantitative shifts for $TT$ and $TE$.
This demonstrates that the MAP lensing estimator delenses in a qualitatively similar way to the quadratic approximations.
Given the comparatively high cost of the MAP method, and the small expected improvements, we chose not to perform the analysis on the curved sky Planck data and simulations.



\twocolumngrid




\allowdisplaybreaks


\bibliography{antony,cosmomc,julien}

\end{document}